\documentclass[10pt]{iopart}
\usepackage[OT1,T1]{fontenc}
\usepackage[latin1]{inputenc}
\usepackage{iopams} %IOP-Paket
\usepackage{setstack}
\usepackage{bbm} %Funktionaldifferential
\usepackage{undertilde} %Untertilde
\newcommand{\di}{\mathrm{d}} %Differential
\newcommand{\ou}[3]{{#1}{}^{#2}{}_{#3}} %Indexstellung
\newcommand{\uo}[3]{{#1}{}_{#2}{}^{#3}} %Indexstellung
\newcommand{\I}{\mathrm{i}} %imaginaere Einheit
\newcommand{\E}{\mathrm{e}} %Euler Zahl
\newcommand{\ellp}{{\ell_{\mathrm{P}}}} %Planck Laenge
\newcommand{\CC}{\mathrm{cc.}} % komplex konjugiertes
\usepackage{dsfont}
\usepackage[dvips]{graphicx}
\newcommand{\LX}{\raisebox{-0.08cm}{\includegraphics[height=2.4ex]{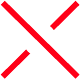}}}
\newcommand{\LZ}{\raisebox{-0.08cm}{\includegraphics[height=2.4ex]{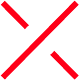}}}
\newcommand{\Lii}{\raisebox{-0.08cm}{\includegraphics[height=2.4ex]{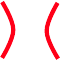}}}
\newcommand{\Lun}{\raisebox{-0.08cm}{\includegraphics[height=2.4ex]{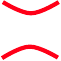}}}
\newcommand{\Lo}{\raisebox{-0.08cm}{\includegraphics[height=2.4ex]{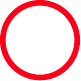}}}
\newcommand{\RX}{\raisebox{-0.08cm}{\includegraphics[height=2.4ex]{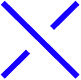}}}
\newcommand{\RZ}{\raisebox{-0.08cm}{\includegraphics[height=2.4ex]{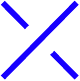}}}
\newcommand{\Rii}{\raisebox{-0.08cm}{\includegraphics[height=2.4ex]{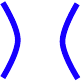}}}
\newcommand{\Run}{\raisebox{-0.08cm}{\includegraphics[height=2.4ex]{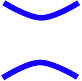}}}
\newcommand{\Ro}{\raisebox{-0.08cm}{\includegraphics[height=2.4ex]{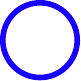}}}
\newcommand{\mixii}{\raisebox{-0.08cm}{\includegraphics[height=2.4ex]{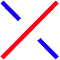}}}
\newcommand{\mixi}{\raisebox{-0.08cm}{\includegraphics[height=2.4ex]{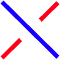}}}
\newcommand{\knot}{\raisebox{-0.08cm}{\includegraphics[height=2.4ex]{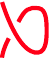}}}
\newcommand{\Li}{\raisebox{-0.08cm}{\includegraphics[height=2.4ex]{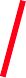}}}
\newcommand{\Lachtv}{\raisebox{-0.08cm}{\includegraphics[height=2.8ex]{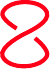}}}
\newcommand{\Lachth}{\raisebox{-0.08cm}{\includegraphics[height=2.3ex]{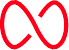}}}
\newcommand{\Linf}{\raisebox{-0.2cm}{\includegraphics[height=3.8ex]{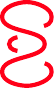}}}
\usepackage{color}
\usepackage{graphicx}
\usepackage{picins}
\newcommand{\qq}[1]{``#1''} %Anfuehrungszeichen
\usepackage{hyperref}
\begin{document}
%%%%%%%%%%%%%%%%%%%%%
%Titel
%%%%%%%%%%%%%%%%%%%%%
\title[Complex Ashtekar variables, the Kodama state and spinfoam gravity]{Complex Ashtekar variables, the Kodama state and spinfoam gravity}

\author{Wolfgang M. Wieland}
\address{Centre de Physique Théorique,\\
   Campus de Luminy, Case 907\\
   13288 Marseille, France, EU}
\eads{%${}^1$\mailto{}, 
\mailto{Wolfgang.Wieland@cpt.univ-mrs.fr}}
\date{April 2010}
\begin{abstract}
Starting from a Hamiltonian description of
four dimensional general relativity in presence of a cosmological constant we perform the program of canonical quantisation. This is done using complex Ashtekar variables while keeping the Barbero--Immirzi parameter real. 
Introducing the $SL(2,\mathbb{C})$ Kodama state formally solving all first class constraints we propose a spinfoam vertex amplitude. 
We construct $SL(2,\mathbb{C})$ boundary spinnetwork functions coloured by finite dimensional representations of the group, and derive the skein relations needed to calculate the amplitude. The space of boundary states is shown to carry a representation of the holonomy flux algebra and can naturally be equipped with a non-degenerate inner product. It fails to be positive definite, but cylindrical consistency is perfectly satisfied. 
\end{abstract}
%%%%%%%%%%%%%%%%%%%%%
%Einleitung
%%%%%%%%%%%%%%%%%%%%%
\section{Introduction}
\paragraph{Motivation and Outline.}
When Hamiltonian general relativity was first formulated \cite{newvariables, ashtekar} in terms of selfdual---i.e. complex valued Ashtekar variables it was the enormous simplification of the Wheeler--DeWitt constraint equation which started all of loop quantum gravity. For a number of technical reasons, the chief of which concerns the construction of the Hilbert space, the complex $SL(2,\mathbb{C})$ variables have later been abandoned in favour of the $SU(2)$ Ashtekar--Barbero \cite{status, thiemann} connection ${}^{(\beta)}\ou{A}{i}{a}=\ou{\Gamma[e]}{i}{a}+\beta\ou{K}{i}{a}$ and its momentum conjugate $\uo{E}{i}{a}=\sqrt{\det h}\uo{e}{i}{a}$. Destroying the polynomial structure of the Hamiltonian constraint equation, this change of variables led to an increasingly more complicated expression for the Wheeler--DeWitt equation.

In this article we want develop arguments, both of kinematical and dynamical nature, which, if further supported by subsequent analysis, may well prove this shift unnecessary. This is done in a number of steps. In section \ref{sect2} we shortly repeat the Hamiltonian formulation of general relativity in presence of a cosmological constant. We use complex Ashtekar variables while keeping the Barbero--Immirzi parameter, unique to all modern versions of loop quantum gravity, real. Compared to the case of $SU(2)$ Ashtekar variables the phase space's dimension increases by a factor of $2\times 2=4$, therefore additional reality conditions need to be imposed. These second class constraint equations decompose into two parts, one of them matching \cite{komplex1} the linear simplicity constraints of spinfoam gravity, the other forcing torsion to vanish. 

Next, we observe that using complex phase space variables the simplification of the Hamiltonian constraint equation is preserved. But having kept the Barbero--Immirzi parameter real, both the selfdual and antiselfdual sectors contribute to the definition of the Hamiltonian constraint. Following \cite{smolinlect, kodama2, GenKod1} after having fixed a particular ordering of the Wheeler--DeWitt constraint equation, we are able to construct a formal solution thereof, i.e. the $SL(2,\mathbb{C})$ Kodama state. Even though all first class constraints are satisfied, the reality conditions are not; this in turn questions the physical significance of the Kodama state. Well aware of the mathematical peculiarities associated to this functional \cite{thiemann, WittenKod},  we certainly not claim this solution describes any proper quantum state of gravity, but use it in order to propose a vertex amplitude for spinfoam gravity. 

In section \ref{sect3} we define $SL(2,\mathbb{C})$ spinnetwork functions on the boundary. On each link, these are coloured by finite dimensional representations of the $SL(2,\mathbb{C})$ group of local Lorentz transformations. They carry a natural representation of the holonomy flux algebra, but do not form a Hilbert space. Abandoning the requirement of positive definiteness while keeping cylindrical consistency, we can then introduce a non degenerate, but indefinite inner product on the space of boundary networks. States can have positive, negative and vanishing norm.

Section \ref{sect4} defines the spinfoam vertex amplitude, essentially being the Chern--Simons expectation value of the $SL(2,\mathbb{C})$ boundary spinnetwork functions, and derives the skein relations needed to calculate it. This follows the analysis of Fröhlich and King \cite{FroehlichCS} generalised to the case of complex gauge group \cite{Wittenkomplex}.

In section \ref{sect5} we propose a strategy in order to implement the reality conditions. We first show that for the boundary states considered this is impossible to achieve in the naive way. In fact Ding and Rovelli implicitly showed in e.g. \cite{physbound} that the reality conditions can only be solved when using unitary (infinite dimensional) representations of the $SL(2,\mathbb{C})$ group. But there may be a way out of this restriction: It is commonly known \cite{GuadagniniCS, 3dimlqg} that Chern--Simons theory naturally leads to a \qq{quantum deformation} of the underlying gauge group. Be that the case for $SL(2,\mathbb{C})$, the reality conditions may get $q$-deformed too. We will then argue that the quantum deformed equations could ultimately be solved by the spinnetwork functions proposed, that is the $q$-deformation may allow us to stick with finite dimensional (chiral) representations of the $SL(2,\mathbb{C})$ group. 

%The major message of the paper is this: Using complex phase space variables, we construct the Kodama state formally solving all first class constraints. This in turns leads to a deformation of the $SL(2,\mathbb{C})$ gauge group, which may well lead us to solve the simplicity constraints by using finite dimensional (chiral) representations of the group.

\paragraph{Notation and conventions.} We adopt the conventions of \cite{komplex1}, abstract indices in (co-)tangent space are labelled by lowercase letters $a,b,c,\dots$ from the beginning of the alphabet, internal indices are denoted by capitals $I,J,K,\dots\in\{0,\dots,3\}$ when referring to four dimensional Minkowski space, or by their lowercase equivalents $i,j,k,\dots\in\{1,2,3\}$ if they correspond to a spatial section thereof. Left handed spinor fields are attached with greek indices $\mu,\nu,\rho\dots$ whereas their right handed counterparts are labelled with \qq{bar-ed} indices $\bar\mu,\bar\nu,\bar\rho\dots$. On $\mathfrak{sl}(2,\mathbb{C})$ we fix an internal time direction and introduce complex coordinates, i.e. we write $\omega=\omega^i\tau_i\in\mathfrak{sl}(2,\mathbb{C})$ where $\omega^i$ is complex and $\sigma_i=2\I\tau_i$ are the Pauli spin matrices.
Indices $I,J,K,\dots\in\{0,\dots,3\}$ are moved by the Minkowski metric $\eta_{IJ}=\mathrm{diag}(-1,1,1,1)_{IJ}$, and the requirement $\epsilon_{0123}=1$ fixes the Levi-Civita pseudotensor in internal space. On a spatial slice $\Sigma$ we write $\tilde{\eta}^{abc}$ for the three dimensional Levi-Civita tensor density, while $\utilde{\eta}_{abc}$ stands for its inverse.

\section{Kodama state for complex Ashtekar variables}\label{sect2}
\subsection{Canonical analysis for complex variables}
Following the program of loop quantum gravity, we choose the Holst action \cite{holst, Parviol} in presence of a cosmological constant $\Lambda$ as the starting point, that is we introduce
\begin{equation}
\eqalign{
S[e,\omega]=\frac{\hbar}{4\ellp^2}\int_Me^I\wedge e^J\wedge&\Big(\epsilon_{IJKL}R^{KL}[\omega]-\frac{2}{\beta}R_{IJ}[\omega]+\\
&-\frac{\Lambda}{6}\epsilon_{IJKL}e^K\wedge e^L
\Big).\label{Hactn}
}
\end{equation}
Where $e$ denotes the co-tetrad, $\omega$ is the $\mathfrak{so}(1,3)$ valued spin connection, and $R[\omega]$ the curvature associated. Apart from $\Lambda$, there are more constants appearing: $\beta\in\mathbb{R}$ is the Barbero--Immirzi parameter and $\ellp=\sqrt{8\pi\hbar G/c^3}$ denotes the rescaled Planck length.

Interested to formulate the theory in the original  \qq{new variables} introduced by Ashtekar \cite{newvariables, ashtekar} let us first decompose the action into its $SO(1,3)$ irreducible parts. This is achieved by means of the selfdual projector $P_{IJMN}$ explicitly introduced later \eref{selfproj}, leading to the expression
\begin{equation}
S[\eta,\omega]=-\frac{\hbar}{2\ellp^2}\frac{\I+\beta}{\I\beta}S_{\mathbb{C}}-\frac{\hbar}{2\ellp^2}\frac{\I-\beta}{\I\beta}\bar{S}_{\mathbb{C}}.\label{actdec}
\end{equation}
Where the complex valued action
\begin{equation}
\eqalign{
S_{\mathbb{C}}&=\int_MP_{IJMN}e^I\wedge e^J\wedge\Big(R^{MN}[\omega]-\frac{\Lambda}{6}e^M\wedge e^N\Big)=\\
&=\int_{\mathbb{R}}\di t\int_\Sigma\Big(-\uo{E}{i}{a}\big(\mathcal{L}_t\ou{A}{i}{a}-D_a\Lambda^i\big)+N^a\ou{F}{i}{ab}\uo{E}{i}{b}+\\
&\qquad\qquad+\frac{\I}{2}\utilde{N}\big(\uo{\epsilon}{i}{lm}\ou{F}{i}{ab}+\frac{\Lambda}{3}\epsilon^{ilm}\utilde{\eta}_{abc}\uo{E}{i}{a}\big)\uo{E}{l}{b}\uo{E}{m}{c}\Big),\label{compactn}
}
\end{equation}
which is the one originally used in loop quantum gravity, naturally appears. In the last two lines of this equation, we performed a $3+1$ decomposition of $M\simeq\mathbb{R}\times\Sigma$, and used a partial gauge fixing (i.e. time gauge) aligning $\eta^0$ to the surface normal of the spatial hypersurface $\Sigma$. The variables $\utilde{N}$ (densitised lapse of weight minus one), $N^a$ (shift vector) and $\Lambda^i=(\frac{1}{2}\uo{\epsilon}{l}{im}\ou{\omega}{l}{ma}+\I\ou{\omega}{i}{oa})t^a$ (where $t^a$ is the time-flow vectorfield transversal to all $t=\mathrm{const.}$ hypersurfaces) are Lagrangian multipliers, all the dynamics is in the Ashtekar variables consisting of
\numparts
\begin{eqnarray}
\mbox{the connection:}&\ou{A}{i}{a}=\ou{\Gamma}{i}{a}+\I\ou{K}{i}{a},\;\mbox{and}\\
\mbox{the densitised triad:}\quad&\uo{E}{i}{a}=\frac{1}{2}\tilde{\eta}^{abc}\epsilon_{ilm}\ou{e}{l}{b}\ou{e}{m}{c}.
\end{eqnarray}   
\endnumparts
Here, and in equation \eref{compactn} above, $\ou{A}{i}{a}$ denotes the $SL(2,\mathbb{C})$ Ashtekar connection, i.e. the spatial projection of the selfdual part of the four dimensional spin connection, $D_a=\partial_a+[A_a,\cdot]$ is the covariant derivative associated, and $\ou{F}{i}{ab}$ is the corresponding curvature, which happens to be the spatial projection of the selfdual part of $R[\omega]$. 

We can then put the equations of motion for $A$, which is complex valued, and $E$ being real, into Hamiltonian form. 
Using the framework of complex variables, this task was---for the case of a vanishing cosmological constant---completely achieved in the paper \cite{komplex1}. The results can immediately be generalised to the presence of a cosmological constant. For the purpose of this article the details of that analysis are of little importance, and we just repeat the main results.
%affine space $\mathcal{A}_{SL(2,\mathbb{C})}$ of smooth 

In order to obtain a Hamiltonian description, let us first introduce the natural symplectic structure on the phase space formed by smooth $SL(2,\mathbb{C})$ connections $\ou{A}{i}{a}$ and corresponding momenta $\uo{\Pi}{i}{a}$ (these are $\mathfrak{sl}(2,\mathbb{C})$ valued vector densities transforming in the adjoint representation of the group) determined by the only non-vanishing Poisson brackets 
\begin{equation}
\eqalign{&\big\{\uo{\Pi}{i}{a}(p),\ou{A}{j}{b}(q)\big\}=\delta^j_i\delta^a_b\delta^{(3)}(p,q),\\
&\big\{\uo{\bar{\Pi}}{i}{a}(p),\ou{\bar{A}}{j}{b}(q)\big\}=\delta^j_i\delta^a_b\delta^{(3)}(p,q).}\label{poissklamm}
\end{equation}
The momentum is related to the densitised triad by
\begin{equation}
\uo{\Pi}{i}{a}\big|_{\mathrm{EOM}}=\frac{\hbar}{2\ellp^2}\frac{\beta+\I}{\I\beta}\uo{E}{i}{a},\quad
\uo{\bar{\Pi}}{i}{a}\big|_{\mathrm{EOM}}=-\frac{\hbar}{2\ellp^2}\frac{\beta-\I}{\I\beta}\uo{E}{i}{a},\label{cmpxcons}
\end{equation}
which can be read from the action (\ref{actdec}, \ref{compactn}). The momentum can a priori take any complex value on phase space, but $\uo{E}{i}{a}$ is necessarily real, i.e. $\mathfrak{su}(2)$ valued, therefore we have to impose the reality conditions
\begin{equation}
\uo{C}{i}{a}=\frac{\ellp^2}{\I\hbar}\Big(\frac{\I\beta}{\beta+\I}\uo{\Pi}{i}{a}+\frac{\I\beta}{\beta-\I}\uo{\bar\Pi}{i}{a}\Big)=0,\label{realcond}
\end{equation}
unexpectedly matching \cite{komplex1} the linear simplicity constraints of spinfoam gravity. These constraints need to be preserved in time, which is possible only if additional secondary constraints are imposed. They require \cite{komplex1} that the spatial part of torsion vanishes:
\begin{equation}
D e+\bar{D} e=2\big(\di e+[\Gamma,e]\big)=0.\label{tors}
\end{equation}
Here $e\equiv\ou{e}{i}{a}$ denotes the $\mathfrak{su}(2)$ valued co-triad. 

Next, we define the densitised triad $\uo{E}{i}{a}$ on the entire complex valued phase space. This is naturally achieved by means of equations (\ref{cmpxcons}, \ref{realcond}) according to
\begin{equation}
\uo{E}{i}{a}=\frac{\ellp^2}{\hbar}\Big(\frac{\I\beta}{\beta+\I}\uo{\Pi}{i}{a}-\frac{\I\beta}{\beta-\I}\uo{\bar\Pi}{i}{a}\Big)=0.\label{Edef}
\end{equation}
Dynamics is determined by a Hamiltonian; like in any reparametrisation invariant theory it is constrained to vanish. It decomposes into a sum over three distinguished terms, Gau\ss, vector and Hamiltonian constraint are explicitly given by:
\numparts
\begin{eqnarray}
G_i[\Lambda^i]&  = & -\int_\Sigma\big(\Lambda^iD_a\uo{\Pi}{i}{a}+\CC\big),\\
H_a[N^a] & = & \int_\Sigma\big(N^a\ou{F}{i}{ab}\uo{\Pi}{i}{b}+\CC\big),\\
\nonumber H[\utilde{N}] & = & -\frac{\ellp^2}{\hbar}\int_\Sigma\utilde{N}\Big(\frac{\beta}{\beta+\I}\uo{\epsilon}{i}{lm}\ou{F}{i}{ab}\uo{\Pi}{l}{a}\uo{\Pi}{m}{b}+\\
&   & \quad+\frac{2\Lambda\ellp^2}{3\hbar}\frac{\I\beta^2}{(\beta+\I)^2}\epsilon^{ilm}\utilde{\eta}_{abc}\uo{\Pi}{i}{a}\uo{\Pi}{l}{b}\uo{\Pi}{m}{c}+\CC\Big).
\end{eqnarray}
\endnumparts
Here the symbol $\CC$ denotes complex conjugation of everything preceding. Both Gau\ss\;and vector constraint have an immediate physical interpretation, the latter generates spatial diffeomorphisms modulo internal gauge transformations whereas the former is associated to pure $SL(2,\mathbb{C})$  transformations. The Hamiltonian constraint is related to spatio-temporal diffeomorphisms along the surface normal.
\subsection{Excursus: The SL(2,C) Chern--Simons functional revisited}
This section gives a short review of the elementary properties of the Chern--Simons functional crucial for the understanding of later parts of this article.
\paragraph{Elementary properties.} Consider the Chern--Simons functional for the $SL(2,\mathbb{C})$ connection, that is the quantity
\begin{equation}
Y[A]=\int_{\Sigma}\mathrm{Tr}\big(A\wedge\di A+\frac{2}{3}A\wedge A \wedge A\big).\label{CSdef}
\end{equation}
We calculate the functional differential and find
\begin{equation}
\mathbbm{d}Y[A]=-\int_{\partial\Sigma}\mathrm{Tr}\big(A\wedge\mathbbm{d}A\big)+2\int_\Sigma\mathrm{Tr}\big(\mathbbm{d}A\wedge F\big).
\end{equation}
If $\Sigma$ is closed the boundary term vanishes and we immediately find that the Chern--Simons functional is a generating potential for the \qq{magnetic} field, i.e.
\begin{equation}
\frac{\delta Y[A]}{\delta \ou{A}{i}{a}(p)}=-\frac{1}{2}\tilde{\eta}^{abc}F_{ibc}\Big|_p\equiv-\uo{B}{i}{a}(p).\label{CSpot}
\end{equation}
In order to obtain the same result for the case of $\Sigma$ being open appropriate boundary conditions need to be imposed (e.g. for  $\Sigma=\mathbb{R}^3$, that $\lim_{r\rightarrow\infty}\iota_r^\ast\mathbbm{d}A=0$, where $\iota_r$ denotes the canonical embedding of the $|\vec{x}|=r=\mathrm{const}.$ 2-sphere into $\mathbb{R}^3$).
\paragraph{Gauge invariance.} 
Since the integrand of \eref{CSdef} manifestly breaks gauge invariance, one may wonder what happens to the whole functional if a gauge transformation is performed. Let $g:SL(2,\mathbb{C})\rightarrow\Sigma$ be a gauge element and consider the transformed connection
\begin{equation}
A^g=g^{-1}\di g+g^{-1}Ag.\label{gaugetrafo}
\end{equation} 
After having performed a partial integration we find
\begin{equation}
\eqalign{
Y[A^g]=Y[A]&-\int_{\partial\Sigma}\mathrm{Tr}\big(\di gg^{-1}\wedge A\big)+\\
&-\frac{1}{3}\int_\Sigma\mathrm{Tr}\big(g^{-1}\di g\wedge g^{-1}\di g\wedge g^{-1}\di g\big).
}
\end{equation}
If we restrict ourselves either to the case of $\Sigma\simeq S_3$ or demand for e.g. $\Sigma\simeq\mathbb{R}^3$ appropriate boundary conditions on $g$, that is $\lim_{r\rightarrow\infty}\iota_r^\ast\di g=0$ where $\iota_r$ again denotes the embedding of the $|\vec{x}|=r=\mathrm{const}.$ 2-sphere into $\mathbb{R}^3$, then the difference of $Y[A^g]-Y[A]$ happens to be proportional to the Brouwer degree
\begin{equation}
n(g):=\frac{1}{24\pi^2}\int_\Sigma\mathrm{Tr}\big(g^{-1}\di g\wedge g^{-1}\di g\wedge g^{-1}\di g\big)
\end{equation}
of $g$, which is a topological invariant taking values in the set of natural numbers \cite{GeometryAndPhysics}. Under these conditions (implicitly assumed in the following) the Chern--Simons functional transforms almost homogeneously under gauge transformations, that is:
\begin{equation}
Y[A^g]-Y[A]=-8\pi^2n(g),\quad n(g)\in\mathbb{Z}.\label{CStrafo}
\end{equation}
Notice also that $n(g)$ is non-vanishing only if the map $g$ cannot continuously be deformed to the identity. Since there is the complex conjugate connection $\ou{\bar{A}}{i}{a}=\ou{\Gamma}{i}{a}-\I\ou{K}{i}{a}$ too, let us also mention that
\begin{equation}
Y[\bar{A}^g]-Y[\bar{A}]=-8\pi^2n(\bar{g})=-8\pi^2\overline{n(g)}=-8\pi^2n(g).
\end{equation}
Where $\bar{g}$ equals the gauge transformation in the complex conjugate representation of the group, i.e. the right handed representation.
\subsection{Kodama state for spinfoam gravity}
Following the general ideas of \cite{kodama1, kodama2, smolinlect, GenKod1} we are now going to construct the Kodama state for chiral variables. This is a \emph{formal} solution of all \emph{first class} Dirac constraints of quantum gravity. Because it lacks a clear physical interpretation this functional is generally not believed \cite{thiemann, WittenKod} to be a proper quantum state of gravity. Nevertheless it may still appear in the definition of the \emph{dynamics} of the theory.

From the canonical Poisson commutation relations \eref{poissklamm} let us first deduce the formal quantisation of the momentum variables
\begin{equation}\eqalign{
\uo{\Pi}{i}{a}(p)\stackrel{\mathrm{q-ation}}{\longrightarrow}\frac{\hbar}{\I}\frac{\delta}{\delta \ou{A}{i}{a}(p)},\\
\uo{\bar\Pi}{i}{a}(p)\stackrel{\mathrm{q-ation}}{\longrightarrow}\frac{\hbar}{\I}\frac{\delta}{\delta \ou{\bar{A}}{i}{a}(p)}.
}
\end{equation}
Choosing an ordering, we are led to the formal quantisation
\begin{equation}
\fl\quad\eqalign{
\widehat{H}[\utilde{N}]  := \ellp^2\hbar\int_\Sigma&\utilde{N}\bigg(\frac{\beta}{\beta+\I}\epsilon^{ilm}\utilde{\eta}_{abc}\frac{\delta}{\delta\ou{A}{l}{b}}\frac{\delta}{\delta\ou{A}{m}{c}}
\Big(\uo{B}{i}{a}+\frac{2\Lambda\ellp^2}{3}\frac{\beta}{\beta+\I}\frac{\delta}{\delta\ou{A}{i}{a}}\Big)+\\
&+\frac{\beta}{\beta-\I}\epsilon^{ilm}\utilde{\eta}_{abc}\frac{\delta}{\delta\ou{\bar{A}}{l}{b}}\frac{\delta}{\delta\ou{\bar{A}}{m}{c}}
\Big(\uo{\bar{B}}{i}{a}-\frac{2\Lambda\ellp^2}{3}\frac{\beta}{\beta-\I}\frac{\delta}{\delta\ou{\bar{A}}{i}{a}}\Big)\bigg)
}
\end{equation}
of the Hamiltonian constraint.
Here $\uo{B}{i}{a}$ denotes the magnetic field, implicitly defined in \eref{CSpot}. Since the Chern--Simons functional is a generating functional for $B$, we can observe that the generalised Kodama state
\begin{equation}
\Omega[A]=\exp\Big(+\frac{3}{2\Lambda\ellp^2}\frac{\beta+\I}{\beta}Y[A]-\frac{3}{2\Lambda\ellp^2}\frac{\beta-\I}{\beta}Y[\bar{A}]\Big)
\end{equation}
is a formal solution of $\widehat{H}[\utilde{N}]\Omega=0$.  Let me emphasise that this state, i.e. the Kodama state generalised to arbitrary values of the Barbero--Immirzi parameter, has originally been derived by Andrew Randono in \cite{GenKod1, GenKod2, GenKod3}.
 
Being diffeomorphism invariant and invariant under small gauge transformation the Kodama state solves both the Gau\ss${}$ and the vector constraint too. 
But the Chern--Simons functional transforms \eref{CStrafo} inhomogeneously under large gauge transformations, and for generic values of $\Lambda$ so does the Kodama state:
\begin{equation}
\eqalign{
\Omega[A^g]&=\Omega[A]\exp\Big(-\frac{12\pi^2}{\Lambda\ellp^2}\frac{\beta+\I}{\beta}n(g)-\CC\Big)=\\
&=\Omega[A]\exp\Big(-\frac{24\I\pi^2}{\Lambda\beta\ellp^2}n(g)\Big).
}
\end{equation}
If we still wish to interpret the Kodama state as a genuine \qq{wave function} it should be single valued, which is possible only if the product $\Lambda\beta\ellp^2$ of the elementary parameters is restricted to discrete values, in other words:
\begin{equation}
\boxed{
\Lambda=\frac{12\pi}{\beta\ellp^2}\frac{1}{n},\quad n\in\mathbb{Z}-\{0\}}\label{lambdadef}
\end{equation}
implying that
\begin{equation}
\Omega[A]=\exp\Big({\frac{n}{8\pi}(\beta+\I)Y[A]-\CC}\Big).\label{Kdmstate}
\end{equation}
Notice the natural appearance of the minimal length scale $\ell_{\mathrm{LQG}}{}^2=\beta\ellp^2$ of loop quantum gravity in the formula for the cosmological constant \eref{lambdadef}. We can also see that the value of $\Omega[A]$ is confined to the unite circle, which suggests that it is not normalisable---at least not with respect to the formal Lebesgue measure $\mathcal{D}[A]$---and therefore lacks to have an interpretation as a proper wave function. Moreover, it does not solve the reality conditions \eref{realcond}, not to mention the difficulties emerging from any attempt to implement the constraints forcing torsion \eref{tors} to vanish. Even if we would gently allow ourselves to ignore all of these mathematical peculiarities, one would have still found just one single solution, and it is far from obvious how one should extract any reasonable physics from it. Of course, there has been research towards this direction \cite{smolinlect}.

However, there may be a way out of this, both mathematically and physically. Using Dirac's bra-ket notation\footnote[7]{In fact we've rather adopted a kind of enhanced Dirac notation commonly used in e.g. \cite{schatten} in order to distinguish between non-normalisable \qq{bras} and hopefully normalisable \qq{kets}.} consider first the Chern--Simons expectation value
\begin{equation}
\left(\Omega|\Psi\rangle\right.=\int_{\mathcal{A}_{SL(2,\mathbb{C})}}\!\!\!\!\!\mathcal{D}[A]\overline{\Omega[A]}\Psi[A]
\end{equation}
of a collection of possibly \emph{intersecting} Wilson lines $\Psi[A]$ evaluated in some irreducible representation of $SL(2,\mathbb{C})$. The integration is performed over the affine space $\mathcal{A}_{SL(2,\mathbb{C})}$ of generalised $SL(2,\mathbb{C})$ connections.  Knowing the results of e.g. \cite{Wittenkomplex, HamquantCS} one may expect that this functional is well defined, and may possess a natural application in quantum gravity. The primary goal of this paper is to argue that within the framework of spinfoam gravity, this is indeed the case. 

Spinfoam models are a kind of path integral quantisation, most of which \cite{lorentzvertam, perez, flppdspinfoam, LQGvertexfinite} can loosely be written as follows
\begin{equation}
Z[\mathcal{C}]=\sum_{\mathrm{colourings}}\!\!\!\!\!\prod_{i=1}^{\#\;\mathrm{of\;faces\;of\;\mathcal{C}}}\!\!\!\!\!\mathsf{A}_{\mathrm{face}}(\rho_i)\!\!\!\!\prod_{j=1}^{\#\;\mathrm{of\;vertices\;of\;}\mathcal{C}}\!\!\!\!\!\mathsf{A}_{\mathrm{vertex}}[\Psi_{v_j}].\label{spnfmmdls}
\end{equation}
Here $Z$ is the partition function associated to a 2-complex $\mathcal{C}$ (a \qq{world-sheet}) consisting of faces, edges and a certain number of vertices. A \qq{colouring} assigns an irreducible representation $\rho_i$ of the gauge group considered to each face, and a normalised intertwiner to each edge. Furthermore, each colouring may be viewed \cite{rovelli} as representing a possible evolution of spinnetwork functions in time. The sum runs over an orthonormal basis of colourings admissible with certain \qq{simplicity constraints} \cite{LQGvertexfinite} matching the reality conditions \eref{realcond} introduced above. 
\begin{figure}[h]
     \centering
     \includegraphics[width= 0.35\textwidth]{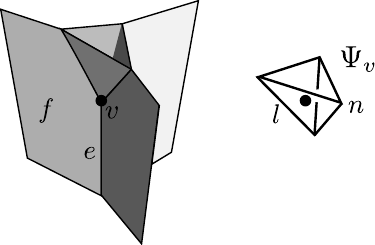}
     \caption{A spinfoam vertex $v$ surrounded by six faces $f$, four internal edges $e$, and the corresponding boundary spinnetwork $\Psi_v$ on the right.}
     \label{spinfoam}
\end{figure}
The face amplitude $\mathsf{A}_{\mathrm{face}}(\rho_i)$ depends only on the colouring of the face (e.g. a half-integer $j_i$ for the case of $SU(2)$), we leave it unspecified here. Dynamics is encoded in the vertex amplitude, which is a functional of the \emph{boundary spinnetwork} $\Psi_v$ obtained \cite{reisenberger} from the intersection of the spinfoam with the boundary of a ball surrounding the vertex $v$. Figure \ref{spinfoam} illustrates this reasoning.

In most spinfoam approaches introduced so far the vertex amplitude evaluates $\Psi$ on flat connections, loosely
\begin{equation}
\mathsf{A}_{\mathrm{vertex}}[\Psi]=\int_{\mathcal{A}_{SL(2,\mathbb{C})}}\!\!\!\!\!\mathcal{D}[A]\delta(F^A)\Psi[A].
\end{equation}
The basic idea we are going to pursue is the following; if we replace in our favourite Lorentzian spinfoam model the vertex amplitude by the Chern-Simons expectation value of the boundary network, that is
\begin{equation}
\mathsf{A}_{\mathrm{vertex}}[\Psi]:=\left(\Omega|\Psi\rangle\right.=\int_{\mathcal{A}_{SL(2,\mathbb{C})}}\!\!\!\!\!\mathcal{D}[A]\E^{-\frac{n}{8\pi}(\beta+\I)Y[A]-\CC}\Psi[A],
\end{equation}
we will manage to find an amplitude in the presence of a cosmological constant \eref{lambdadef} formally solving the Hamiltonian constraint in the sense that:
\begin{equation}
\phantom{)}\big(\Omega\big|\widehat{H}[\utilde{N}]^\ast\Psi\big\rangle=\phantom{)}\big(\widehat{H}[\utilde{N}]\Omega\big|\Psi\big\rangle=0.\label{Sampl}
\end{equation}
---a proposal clearly inspired by the pioneering works of \cite{craneyetter, kosmdeform, Baezartcl1, Baezartcl2}. In order to realise this possibility the following points need to be clarified:
\begin{enumerate}
\item Specify the space of boundary spinnetworks $\Psi$ used, and equip it with a Hilbert space structure. This task will be discussed in section \ref{sect3}.
\item Calculate the Chern--Simons expectation value for the boundary states chosen. Section \ref{sect4} is dedicated to this goal.
\item Restrict the spinnetworks to those solving the reality conditions \eref{realcond} in an appropriate way. At the end of this paper, leaving a detailed analysis open, we argue that certain boundary states may realise this condition.
\end{enumerate} 

\section{Chiral representations for quantum gravity}\label{sect3}
\subsection{Motivation}
In order to do quantum theory, we need an inner product. If we try to repeat for complex variables what has successfully been achieved for the case of $SU(2)$, we quickly run into serious troubles. In fact, the non-compactness of the $SL(2,\mathbb{C})$ group prevents us to use the most natural inner product, i.e the one borrowed from the Haar measure
\begin{equation}
d\mu(g)\propto \mathrm{Tr}\big(g^{-1}\di g\wedge g^{-1}\di g\wedge g^{-1}\di g\big)\wedge\CC\label{haarmeas}
\end{equation}
What are the difficulties with this measure? First, it is unique only up to an overall constant. This has recently been observed \cite{Lorentzambi} to be an ambiguity of all current Lorentzian spinfoam models. The other problem closely related to the former is that this measure associates an infinite volume
\begin{equation}
\int_{SL(2,\mathbb{C})}d\mu(g)=\infty\label{divergence}
\end{equation}
to the group, preventing us to use it for quantum gravity. The basic argument goes like this. The space of cylindrical functions \emph{on a single graph} naturally inherits an inner product from the one chosen in the underlying gauge group. However the inner product so-defined has to fulfil several requirements, the chief of which is cylindrical consistency:

Any cylindrical function is cylindrical with respect to infinitely many different graphs (e.g. cut paths into pieces). Having defined an inner product on each possible graph separately, one has to prove that, given any two cylindrical functions their common inner product is independent of the graph chosen in order to calculate it. In the case of a divergent integral like the one in \eref{divergence} this is impossible to achieve.

In the following we develop an argument against this reasoning. Abandoning the requirement of positive-definiteness we introduce a non-degenerate sesquilinearform (in the following loosely called \qq{inner product}) on $SL(2,\mathbb{C})$ fulfilling all criteria needed to prove cylindrical consistency. By this definition, we encounter states of positive, vanishing and even negative norm on the group. From the sesquilinearform so defined an inner product on the space of cylindrical functions of the $SL(2,\mathbb{C})$ connection will be constructed in the obvious way. It will inherit the indefiniteness of the former, giving raise to the notion of unphysical states of negative or even
vanishing norm. This can be compared with the appearance of negative norm states in the Gupta--Bleuler \cite{gupta, bleuler} formalism, where on the one particle space of both longitudinal and transversal photons there is just an indefinite inner product
\begin{equation}
\langle\varepsilon,\vec{k}|\varepsilon^\prime,\vec{k}^\prime\rangle_{\mathrm{GB}}=\frac{\eta^{IJ}\varepsilon_I\varepsilon^\prime_J}{2|\vec{k}|}\delta^{(3)}(\vec{k}-\vec{k}^\prime)
\end{equation}
available.

In order to develop our arguments let us first recapitulate some basic tools of the representation theory of the Lorentz group. 

\subsection{Prerequisite: Representation theory of the Lorentz group}
Following the analysis of \cite{sexlurbantke2}, we present the most basic facts about the finite dimensional representations of the Lorentz group. 
Consider first the complexified Lie algebra $\mathbb{C}\times\mathfrak{so}(1,3)$ of the Lorentz group and the natural adjoint group action therein. Restricting ourselves to proper orthochronous transformations $L_+^\uparrow\subset SO(1,3)$ this representation turns out to be reducible. In fact if we anticipate some notation being introduced later, it decomposes according to $(1,0)\oplus (0,1)$ into irreducible ones. The maps towards its two irreducible components are achieved by the self- and antiselfdual projectors $P$ and $\bar{P}$ explicitly given by
\begin{equation}
\ou{P}{IJ}{MN}=\frac{1}{2}\big(\delta^{[I}_{M}\delta^{J]}_{N}-\frac{\I}{2}\ou{\epsilon}{IJ}{LM}\big)\label{selfproj}
\end{equation}
and its complex conjugate. A natural basis in the image of $P$ and $\bar{P}$ respectively is given by the self- and antiselfdual generators
\begin{equation}
T_i=\frac{1}{2}(L_i-\I K_i),\quad\bar{T}_i=\frac{1}{2}(L_i+\I K_i),
\end{equation}
constructed from infinitesimal boosts $K_i$ and rotations $L_i$, obeying the algebraic relations
\numparts
\begin{eqnarray}
\left[T_i,T_j\right]&=&\I\uo{\epsilon}{ij}{l}T_l,\label{comm1}\\
\left[T_i,\bar{T}_j\right]&=&0,\label{comm2}\\
\left[\bar{T}_i,\bar{T}_j\right]&=&\I\uo{\epsilon}{ij}{l}\bar{T}_l.\label{comm3}
\end{eqnarray}
\endnumparts
Using this complex basis elements, any $\omega\in\mathfrak{so}(1,3)\simeq\mathfrak{sl}(2,\mathbb{C})$ decomposes according to
\begin{equation}
\omega=-\I\omega^iT_i-\I\bar{\omega}^i\bar{T}_i,
\end{equation}
into complex components $\omega^i\in\mathbb{C}^3$, where the real and imaginary parts of $\omega^i$ correspond to rotations and boosts respectively. Since (\ref{comm1}, \ref{comm2}, \ref{comm3}) are nothing but the commutation relations for two pairs of the rotation group, we can immediately find all finite dimensional irreducible representations of this algebra. In fact, irreducible representations are labelled by a pair of spins, i.e. half integers $(j,k)$. By exponentiation, we obtain the associated representations of the (universal cover of the) Lorentz group, which turn out to be the analytical continuation
\begin{equation}
\ou{\big[D^{(j,k)}(g)\big]}{\mu\bar{\rho}}{\mu\bar{\sigma}}=\ou{\big[D^{(j)}(g)\big]}{\mu}{\nu}\overline{\ou{\big[D^{(k)}(g)\big]}{\rho}{\sigma}}\label{comWig}
\end{equation}
of the Wigner matrices $D^{(j)}$ to the complex. Here the bi-index $(\mu,\bar\mu)$ refers to the $(2j+1)\times(2k+1)$-dimensional representation space
\begin{equation}
(j,k):=\mathrm{Sym}\big(\bigotimes^{2j}\mathbb{C}^2\big)\otimes\mathrm{Sym}\big(\bigotimes^{2k}\bar\mathbb{C}^2\big)\label{repspace}
\end{equation}
constructed from the symmetriced tensor products of $\mathbb{C}^2$ and the complex conjugate vector space $\bar\mathbb{C}^2$. We call an element of $(j,0)$ a left-handed spinor, and for the $(0,j)$ sector of opposite chirality we call them right-handed. Note that in order to relate the complex conjugate vector space to $\mathbb{C}^2$, some basis elements have to be chosen to be \qq{real}. And \qq{bar-ed} indices (e.g. $\bar\mu,\,\bar\nu\dots$) are used for spinors in the complex conjugate representation spaces, rather than the more common primed-index notation (e.g. $\mu^\prime,\,\nu^\prime\dots$). 

The argument $g$ in equation \eref{comWig} lies in $SL(2,\mathbb{C})$, i.e. the universal covering group of the Lorentz group $L_+^\uparrow=SL(2,\mathbb{C})/_{\{-\mathds{1},\mathds{1}\}}$. In fact the relation between the two is provided by the intertwining Pauli matrices $\sigma_\mu=(\mathds{1},\vec{\sigma})$ and the relations
\begin{equation}
\ou{g}{\mu}{\nu}\ou{\bar g}{\bar\mu}{\bar\nu}(\sigma_I)^{\nu\bar\nu}=\ou{\Lambda(g)}{J}{I}(\sigma_J)^{\mu\bar\mu},\;\forall g\in SL(2,\mathbb{C}),\Lambda(g)\in L_+^\uparrow.
\end{equation}
At the level of the Lie algebra, we find that
\begin{equation}
\mbox{for\;} g=\exp(\omega^i\tau_i):\Lambda(g)=\exp\big(-\I\omega^iT_i-\I\bar\omega^i\bar{T}_i\big),
\end{equation}
where $\sigma_i=2\I\tau_i$ denote the Pauli spin matrices.
In each representation space there is the invariant $\epsilon$-tensor, determined by the requirement
\begin{equation}
\forall g\in SL(2,\mathbb{C}):{}^{j}{\epsilon}_{\mu\nu}\ou{[D^{(j,0)}(g)]}{\mu}{\alpha}\ou{[D^{(j,0)}(g)]}{\nu}{\beta}={}^{j}{\epsilon}_{\alpha\beta}.
\end{equation}
In the canonical $|j,m\rangle\in(j,0)$ basis its matrix entries are chosen to be
\begin{equation}
{}^{j}{\epsilon}_{mm^\prime}={}^j{\epsilon}^{mm^\prime}=\overline{{}^{j}{\epsilon}^{mm^\prime}}=(-1)^{m-j}\delta_{m,-m^\prime}.
\end{equation}
Being a tensorial invariant, the $\epsilon$-tensor can naturally be used to rise and lower vector indices in order to define a map between $(j,k)$ and its dual vector space, e.g:
\begin{equation}\eqalign{
v^\alpha\in(j,0)\mapsto v_\alpha={}^j\epsilon_{\beta\alpha}v^\beta&\in(j,0)^\prime\\
\xi_\alpha\in(j,0)^\prime\mapsto\xi^\alpha={}^j\epsilon^{\alpha\beta}\xi_\beta&\in(j,0)
}
\end{equation}
The relation with $SO(4)$ deserves some further attention. For $SO(4)$ irreducible representations are again labelled by a pair $(j,k)_{\mathrm{E}}$ of spins. These representations happen to be \emph{unitary}. Therefore each representation space is equipped with an additional structure, there is an \emph{invariant Hermitian} inner product, i.e. a metric tensor ${}^j\delta_{\mu\bar\mu}$ absent in the Lorentzian case available. This in turn allows to map \qq{bar-ed} indices $\bar\mu\dots$ to \qq{ordinary} indices. And hence, the complex conjugate representation is immediately shown to be homomorphic to the original one $\overline{(j,k)}_{\mathrm{E}}\simeq(j,k)_{\mathrm{E}}$. For the Lorentzian case, the situation is different. Indeed equation \eref{repspace} shows complex conjugation to do nothing but to exchange the representation spaces between one another, i.e.
\begin{equation}
\overline{(j,k)}=(k,j).
\end{equation}
%And we write for complex conjugation:
%\begin{equation}
%\xi^{\rho\bar\mu}\in(j,k)\stackrel{\CC}{\rightarrow}\overline{\xi^{\rho\bar\mu}}\equiv\bar{\xi}^{\mu\bar{\rho}}
%\end{equation}
In particular:
\begin{equation}
\overline{\ou{[D^{(j,k)}]}{\mu\bar\rho}{\nu\bar\sigma}}=\ou{[D^{(k,j)}]}{\rho\bar\mu}{\sigma\bar\nu}.
\end{equation}
\subsection{An inner product on the group}
In this section we are going to introduce an inner product on a suitable space of functions on the $SL(2,\mathbb{C})$ group, fulfilling all requirements needed to use it for the quantum theory.

We start with the space of polynomials. More precisely let us call any $f:SL(2,\mathbb{C})\rightarrow\mathbb{R}$ a polynomial on the group, symbolically denoted by $f\in \mathrm{Poly}_{SL(2,\mathbb{C})}$, provided it decomposes into a finite complex linear combination of monomials $a^m\bar{a}^nb^r\bar b^s\dots$ generated by the matrix elements $a,b\dots$ of group elements and their complex conjugate (and of course $m,n,r,\dots\in\mathbb{N}_0$). Being associated to a finite dimensional product representation of $SL(2,\mathbb{C})$ any of these monomials reduce to a finite sum over irreducible representations. Therefore all of $\mathrm{Poly}_{SL(2,\mathbb{C})}$ is actually already generated by the matrix elements of the finite dimensional representations of the group. Any polynomial on $SL(2,\mathbb{C})$ can thus be decomposed according to
\begin{equation}
f(g)=\sum_{2j=0}^N\sum_{2k=0}^M \sqrt{2j+1}\sqrt{2k+1}\ou{f^{(j,k)}}{\mu\bar\mu}{\nu\bar\nu} \ou{\big[D^{(j,k)}(g)\big]}{\nu\bar\nu}{\mu\bar\mu}.
\end{equation}
Where $M,N<\infty$, and $\ou{f^{(j,k)}}{\mu\bar\mu}{\nu\bar\nu}$ are some spinorial coefficients. On $\mathrm{Poly}_{SL(2,\mathbb{C})}$ there exists a natural $SL(2,\mathbb{C})$ invariant linear functional, entirely fixed by its action
\begin{equation}
\eta:\mathrm{Poly}_{SL(2,\mathbb{C})}\rightarrow\mathbb{C}:\eta\big(\ou{[D^{(j,l)}]}{\nu\bar\nu}{\mu\bar\mu}\big)=\delta_{j,0}\delta_{l,0}.\label{etafunc}
\end{equation}
on the basis elements of the polynomials. This functional fulfils several important properties of invariance. If we introduce left and right translation together with the inverse operation on the space of polynomials on the group, that is
\begin{equation}
\forall g,h\in SL(2,\mathbb{C}), f\in\mathrm{Poly}_{SL(2,\mathbb{C})}:\left\{\eqalign{\big(L_gf\big)(h)&=f(gh)\\\big(R_gf\big)(h)&=f(hg)\\\big(Sf\big)(h)&=f(h^{-1})}\right.
\end{equation}
by equation \eref{etafunc} we realise that
\begin{equation}
\forall g\in SL(2,\mathbb{C}), f\in\mathrm{Poly}_{SL(2,\mathbb{C})}:\left\{\eqalign{\eta(f)&=\eta(L_gf)=\\&=\eta(R_gf)=\eta(Sf).}\right.\label{invar}
\end{equation}
Setting furthermore
\begin{equation}
\forall f,f^\prime\in\mathrm{Poly}_{SL(2,\mathbb{C})}:\langle f,f^\prime\rangle=\eta(\bar{f}f^\prime)\label{innprod}
\end{equation}
we succeeded in introducing a $SL(2,\mathbb{C})$ invariant sesquilinearform on $\mathrm{Poly}_{SL(2,\mathbb{C})}$. In order to simplify our notation it will proof useful to use an integral notation, i.e. we write
\begin{equation}
\eta(\bar{f}f^\prime)=\langle f,f^\prime\rangle\equiv\int_{SL(2,\mathbb{C})} \mathfrak{D}(g)\overline{f(g)}f^\prime(g).\label{measdef}
\end{equation}
Notice that we do not claim to rigorously construct $\mathfrak{D}(g)$ as a proper measure on $SL(2,\mathbb{C})$ or on any sophisticated compactification thereof. At this level, equation \eref{measdef} is nothing but notation, a simple graphical tool later used in order to bring our equations in a form maximally close to those of the compact case of $SU(2)$. Let us give an example here, using this integral notation the properties of \eref{invar} compactly read $\forall f\in\mathrm{Poly}_{SL(2,\mathbb{C})}$ and $h\in SL(2,\mathbb{C})$:
\begin{equation}
\eqalign{
\int\mathfrak{D}(g)f(g)&=\int\mathfrak{D}(g)\int\mathfrak{D}(h)f(gh)=\\
&=\int\mathfrak{D}(g)f(gh)=\int \mathfrak{D}(g)f(hg)=\int \mathfrak{D}(g)f(g^{-1}).}\label{invar2}
\end{equation}
This formal integration has a certain number of interesting features. First of all it associates a finite volume to the group. Indeed
\begin{equation}
\int_{SL(2,\mathbb{C})} \mathfrak{D}(g)=\eta(1)=1.
\end{equation}
Next, it cannot be positive, in fact there are states of negative, e.g.
\begin{equation}
\eqalign{
\big\langle \ou{[D^{({1/2},0)}]}{0}{0}&-\ou{[D^{(0,{1/2})}]}{\bar{1}}{\bar{1}},
\ou{[D^{({1/2},0)}]}{0}{0}-\ou{[D^{(0,{1/2})}]}{\bar{1}}{\bar{1}}\big\rangle=\\
&=-\frac{1}{2}\bar\epsilon^{\bar{0}\bar{1}}\bar\epsilon_{\bar{0}\bar{1}}-\frac{1}{2}\epsilon^{10}\epsilon_{10}=-1,\label{negnorm}
}
\end{equation}
together with those of vanishing, e.g.
\begin{equation}
\big\langle \ou{[D^{(1,0)}]}{1}{0},\ou{[D^{(1,0)}]}{1}{0}\big\rangle=0\label{vannorm},
\end{equation}
norm. Of course there are states of positive norm too, consider $f\in\mathrm{Poly}_{SL(2,\mathbb{C})}$ defined by
\begin{equation}
f(g)=\mathrm{Tr}\big(D^{(j,k)}(g)\big)+\mathrm{Tr}\big(D^{(k,j)}(g)\big).\label{posnorm1}
\end{equation}
Then
\begin{equation}
\langle f,f\rangle=2.\label{posnorm2}
\end{equation}
In full generality, the basic elements of $\mathrm{Poly}_{SL(2,\mathbb{C})}$ fulfil the following \qq{orthogonality} relations
\begin{equation}
\eqalign{
\big\langle &\ou{[D^{(j,k)}]}{\mu\bar\rho}{\nu\bar\sigma},
\ou{[D^{(j^\prime,k^\prime)}]}{\mu^\prime\bar\rho^\prime}{\nu^\prime\bar\sigma^\prime}\big\rangle=\\
&\quad=\frac{1}{2j+1}\frac{1}{2k+1}
\delta_{j,k^\prime}\delta_{j^\prime,k}\,
{}^{k}\epsilon^{\rho\mu^\prime}\,{}^{k}\epsilon_{\sigma\nu^\prime}\,{}^{j}\bar\epsilon^{\bar\mu\bar\rho^\prime}\,{}^{j}\bar\epsilon_{\bar\nu\bar\sigma^\prime}
}
\end{equation}
which implicitly prove non-degeneracy of the inner product $\langle\cdot,\cdot\rangle$ on $\mathrm{Poly}_{SL(2,\mathbb{C})}$. That is
\begin{equation}
\langle f,f^\prime\rangle=0\;\forall f^\prime\in\mathrm{Poly}_{SL(2,\mathbb{C})}\Leftrightarrow f=0.
\end{equation}
The orthogonality relations give rise to the decomposition of the identity, i.e. for all $f\in\mathrm{Poly}_{SL(2,\mathbb{C})}$
\begin{equation}
f(g)=\sum_{j,k\in\mathbb{N}_0}(2j+1)(2k+1)\ou{[D^{(j,k)}(g^{-1})]}{\mu\bar\rho}{\nu\bar\sigma}\big\langle \ou{[D^{(k,j)}]}{\sigma\bar\nu}{\rho\bar\mu},f\rangle.\label{iddecomp}
\end{equation}
If we now define the delta functional
\begin{equation}
\forall f\in\mathrm{Poly}_{SL(2,\mathbb{C})}:\delta(f)=f(\mathds{1})
\end{equation}
on the group, we immediately arrive at the familiar looking character decomposition
\begin{equation}
\delta(\,\cdot\,)=\sum_{j,k\in\frac{1}{2}\mathbb{N}_0}(2j+1)(2k+1)\big\langle \ou{[D^{(k,j)}]}{\mu\bar\nu}{\mu\bar\nu},\,\cdot\,\rangle
\end{equation}
thereof. 
\subsection{An inner product for quantum theory}
In the last section we have introduced an inner product on the group. Indefinite, but fulfilling a number of properties of invariance, we now lift it to the quantum theory. 

In order to do this, let us first introduce the notion of cylindrical functionals of the $SL(2,\mathbb{C})$ Ashtekar connection $\ou{A}{i}{a}=\ou{\Gamma}{i}{a}+\I\ou{K}{i}{a}$. These are constructed in the obvious way as in \cite{thiemann, rovelli}. Consider a graph $\Gamma=(\gamma_1,\dots,\gamma_L)$, i.e. a finite collection of (piecewise analytic) oriented paths (links). Adopting the usual notion from the $SU(2)$ case, we call a functional $\Psi$ of the $SL(2,\mathbb{C})$ Ashtekar connection \emph{cylindrical} with respect to $\Gamma$, symbolically denoted by $\Psi\in\mathrm{Cyl}_\Gamma$ provided
\begin{equation}
\eqalign{
\Psi\in\mathrm{Cyl}_\Gamma\Leftrightarrow\exists f\in\mathrm{Poly}_{SL(2,\mathrm{C})^L}:\\
\forall A\in\mathcal{A}_{SL(2,\mathbb{C})}:\Psi[A]=f\big(h_{\gamma_1}[A],\dots, h_{\gamma_L}[A]\big).}\label{cyldef}
\end{equation}
Where $h_\gamma[A]=\boldsymbol{P}\exp(-\int_\gamma A)\in SL(2,\mathbb{C})$ denotes the holonomy, parallely transporting left handed Weyl spinors all along $\gamma$, $\mathrm{Poly}_{SL(2,\mathbb{C})^L}$ equals the space of polynomials on a number of $L$ copies of $SL(2,\mathbb{C})$, and $\mathcal{A}_{SL(2,\mathbb{C})}$ again denotes the affine space of (generalised) $SL(2,\mathbb{C})$ connections \cite{thiemann}. We may now wish to call $\Psi$ cylindrical (written as $\Psi\in\mathrm{Cyl}$) provided there exists some graph $\Gamma$ such that $\Psi\in\mathrm{Cyl}_\Gamma$. By this very definition we first notice that any element of $\mathrm{Cyl}$ is necessarily cylindrical with respect to infinitely many different graphs (those constructed from inversion of the orientation of links, the introducton of links coloured by the trivial representation, and those emerging if links are cut into finite pieces). 

Everything has been collected to equip $\mathrm{Cyl}_\Gamma$ with an inner product borrowed from the one introduced in \eref{innprod}. Given any two cylindrical functions $\Psi_f,\Psi_{f^\prime}\in\mathrm{Cyl}_\Gamma$ being cylindrical with respect to the very same graph $\Gamma$, such that $\Psi_f[A]=f(h_{\gamma_1}[A],\dots)$ and equally for $f^\prime$, we define their common inner product to be
\begin{equation}
\big\langle\Psi_f,\Psi_{f^\prime}\big\rangle_{\Gamma}:=\eta(\bar{f}f^\prime)=\langle f,f^\prime\rangle.
\end{equation}
Where we implicitly extended $\eta$ to $\mathrm{Poly}_{SL(2,\mathbb{C})\times SL(2,\mathbb{C})\times\dots}$. This is done in the obvious way; for $f_1, f_2,\dots\in\mathrm{Poly}_{SL(2,\mathbb{C})}$ the equation $(f_1\times f_2\times\dots)(g_1,g_2,\dots)=f_1(g_1)f_2(g_2)\cdots$ defines an element
$(f_1\times f_2\times\dots)\in\mathrm{Poly}_{SL(2,\mathbb{C})\times SL(2,\mathbb{C})\times\dots}$.
We now set $\eta(f_1\times f_2\times\dots)=\eta(f_1)\eta(f_2)\cdots$, which by linearity extends $\eta$ to all of $\mathrm{Poly}_{SL(2,\mathbb{C})\times{SL(2,\mathbb{C})}\times\dots}$. By this very definition the inner product $\langle\,\cdot\,,\,\cdot\,\rangle$ naturally inherits the non-degeneracy from the one chosen on the group, that is:
\begin{equation}
\big\langle\Psi,\Psi^\prime\big\rangle_\Gamma\;\forall\Psi^\prime\in\mathrm{Cyl}_\Gamma\Leftrightarrow\Psi=0
\end{equation}
Having defined the inner product on each different graph separately, we are facing the question of cylindrical consistency. In other words, can the inner product so-defined be extended to all of $\mathrm{Cyl}$? 
\begin{figure}[h]
     \centering
     \includegraphics[width= 0.4\textwidth]{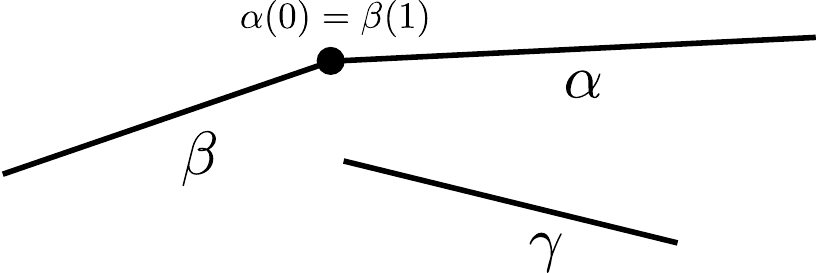}
     \caption{On cylindrical consistency.}
     \label{cons}
\end{figure}
Let's discuss the problem associated in a reduced example. Consider three links $\alpha$, $\beta$ and $\gamma$, suppose $\alpha\cap\gamma=\beta\cap\gamma=\emptyset$ and $\alpha(0)=\beta(1)$ (see figure \ref{cons} for an illustration). Take some $f,f^\prime\in \mathrm{Poly}_{SL(2,\mathbb{C})}$. Construct $\Psi[A]=f\big(h_{\alpha\circ\beta}[A]\big)$ and $\Psi^\prime[A]=f^\prime\big(h_{\alpha\circ\beta}[A]\big)$. Thus both $\Psi,\Psi^\prime\in\mathrm{Cyl}$. A short moment of reflection reveals that both $\Psi$ and $\Psi^\prime$ are cylindrical with respect to the very different graphs $(\alpha\circ\beta)$, $(\alpha,\beta,\gamma)$, $(\alpha^{-1},\beta,\gamma^{-1})$. However by virtue of the properties \eref{invar2} of $\eta$, for all these mutually different graphs the value of the inner product turns out to be the same:
\begin{equation}
\langle\Psi,\Psi^\prime\rangle_{(\alpha\circ\beta)}=\langle\Psi,\Psi^\prime\rangle_{(\alpha,\beta,\gamma)}=\langle\Psi,\Psi^\prime\rangle_{(\alpha^{-1},\beta,\gamma^{-1})}
\end{equation} 
This observation holds for the general case, allowing us to extend the inner product to all of $\mathrm{Cyl}$: Given two elements $\Psi,\Psi^\prime$, we can always find a graph $\Gamma$ sufficiently large such that $\Psi,\Psi^\prime\in\mathrm{Cyl}_\Gamma$. We then define their common inner product to be the one associated to $\Gamma$, i.e.
\begin{equation}
\langle\Psi,\Psi^\prime\rangle_{\mathrm{Cyl}}:=\langle\Psi,\Psi^\prime\rangle_\Gamma,
\end{equation}
and again by the use of the elementary properties of $\eta$, we find the value of the inner product to be independent of the graph chosen. In other words, we have equipped $\mathrm{Cyl}$ with a natural non-degenerate inner product, i.e. one being cylindrical consistent.
 
Let us again stress the fact that this would have been impossible to achieve if we had tried to use an inner product borrowed from the usual positive-definite Haar measure \eref{haarmeas}. The divergent volume of the group $\int_{SL(2,\mathbb{C})}d\mu(g)=\infty$ preventing to achieve cylindrical consistency would have made any such effort doomed to fail.  

Having introduced an inner product we do not yet have a Hilbert space. Even worse, all the peculiarities we encountered with the indefinite inner product on $SL(2,\mathbb{C})$ are now inherited by $\langle\cdot,\cdot\rangle_{\mathrm{Cyl}}$. In fact from \eref{negnorm}, and \eref{vannorm} we can straight forwardly construct elements of $\mathrm{Cyl}$ having negative or even vanishing norm.  

However there certainly are elements of $\mathrm{Cyl}$ equipped with a positive norm. Let us give an example here. Remember first equation \eref{posnorm1}, consider now some loop $\alpha$. Define
\begin{equation}
\Psi_\alpha[A]=\mathrm{Tr}\big(D^{(j,k)}(h_\alpha[A])\big)+\mathrm{Tr}\big(D^{(k,j)}(h_\alpha[A])\big)\label{firstexam}
\end{equation}
in order to observe \eref{posnorm2} positivity
\begin{equation}
\big\langle\Psi_\alpha,\Psi_\alpha\big\rangle_{\mathrm{Cyl}}=2\label{postvty}
\end{equation}
of the norm. This is an important observation, allowing for immediate generalisation. In fact on the subspace of $\mathrm{Cyl}$ generated by all finite complex linear combinations of finite products of non overlapping (but possibly intersecting) Wilsonian lines of the type of \eref{firstexam}, the inner product will remain positive definite. Far from claiming that this subspace has any physical importance we want to stress that its sheer existence suggests the following:

Suppose the reality conditions \eref{realcond} could be solved in some appropriate way---e.g. in the sense of Gupta and Bleuler---on a subspace $\mathcal{K}$ of $\mathrm{Cyl}$. Provided furthermore that on $\mathcal{K}$ the inner product remains positive (semi-)definite (which by virtue of equation \eref{postvty} remains logically possible) we could immediately turn it into a Hilbert space. On the kinematical level this would be a major step towards the goal of canonical quantisation for \emph{complex variables}.
\subsection{Holonomy flux algebra}
The canonical Poisson commutation relations in the form of equation \eref{poissklamm} behave too singular and cannot directly be implemented on the space of cylindrical functions introduced above. In loop quantum gravity one therefore studies suitably smeared phase space variables instead. These are the holonomy along a link $\gamma$ together with the momentum conjugate smeared over a two dimensional oriented surface $f$. Proving \cite{thiemann} that these smeared quantities (i) can still capture the full phase space and (ii) form a closed algebra under the Poisson bracket, they serve as the starting point for the canonical quantisation program of loop quantum gravity. For the case of complex \cite{ashtekar, newvariables} Ashtekar variables they are
\numparts
\begin{eqnarray}
\mbox{the holonomy:\quad}h_{\gamma}[A]=\boldsymbol{P}\exp\big(-\int_\gamma A\big)\label{holdef}\quad\mbox{and,}\\
\mbox{the flux:\quad}\Pi_i[f]=\int_{f}\utilde{\eta}_{abc}\uo{\Pi}{i}{a},\label{momsmear}
\end{eqnarray}
\endnumparts
together with their complex conjugate. Notice that in \eref{momsmear} the tensor density $\utilde{\eta}_{abc}$ cancels the weight of $\uo{\Pi}{i}{a}$ leaving us with a 2-form (in abstract index notation) which can naturally be integrated over any oriented surface $f$. 
\begin{figure}[h]
     \centering
     \includegraphics[width= 0.25\textwidth]{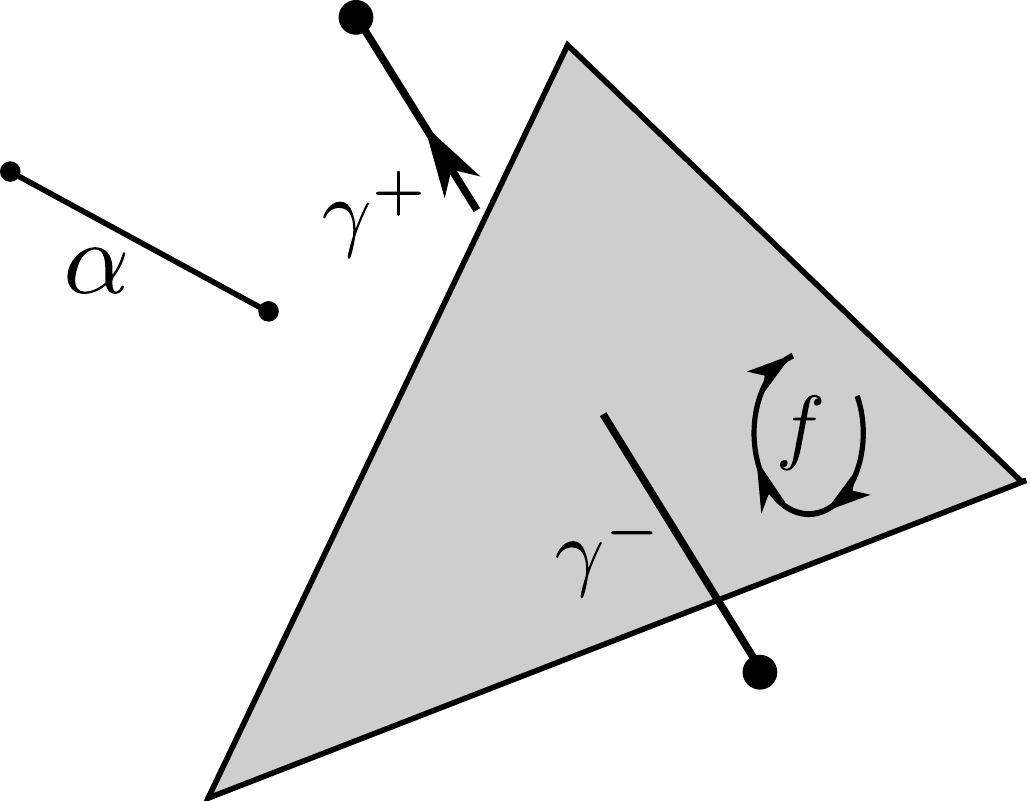}
     \caption{Phase space variables are smeared over links and faces.}
     \label{smearng}
\end{figure}
These smeared variables obey the Poisson algebra of lattice gauge theory, i.e. the holonomy flux algebra of $SL(2,\mathbb{C})$. Consider the following reduced example. Let $f$ be a surface intersecting some link $\gamma$ exactly once, cutting $\gamma$ into pieces $\gamma=\gamma^+\circ\gamma^-$. Let additionally $\alpha$ be another link away from $f$, i.e. $f\cap\alpha=\emptyset$. For the left-handed sector we find:
\numparts
\begin{eqnarray}
\big\{\Pi_i[f],h_\alpha\big\}=\big\{h_\gamma,h_\alpha\big\}=0\label{zerobrace}\\
\big\{\Pi_i[f],h_\gamma\big\}=-\epsilon(f,\gamma)h_{\gamma^+}\tau_ih_{\gamma^-},\label{PiUbrace}%\\
%\big\{\Pi_i[f],\Pi_j[f]\big\}=-\uo{\epsilon}{ij}{k}\Pi_k[f].\label{PiPibrace}
\end{eqnarray}
\endnumparts
We are then left to calculate the right handed sector which is nothing but the complex conjugate of the former:
\numparts
\begin{eqnarray}
\big\{\bar{\Pi}_i[f],\bar{h}_\alpha\big\}=\big\{\bar{h}_\gamma,\bar{h}_\alpha\big\}=0\label{zerobrace2}\\
\big\{\bar\Pi_i[f],\bar{h}_\gamma\big\}=-\epsilon(f,\gamma)\bar{h}_{\gamma^+}\bar{\tau}_i\bar{h}_{\gamma^-},\label{PiUbrace2}%\\
%\big\{\bar\Pi_i[f],\bar\Pi_j[f]\big\}=-\uo{\epsilon}{ij}{k}\bar\Pi_k[f].\label{PiPibrace2}
\end{eqnarray}
\endnumparts
All Poisson brackets between variables of mutually different chirality vanish. 
Here $\epsilon(f,\gamma)$ denotes the relative orientation of $f$ and $\gamma$. Notice furthermore that the Poisson bracket $\{\Pi_i[f],\Pi_j[f]\}$ is missing in this overview, in fact it does not vanish, a property either traced back to the Jacobi identity \cite{noncomm}, or even more elegantly to a smearing \cite{steppingout, EEcomm} of the momentum variables slightly different than the one chosen in \eref{momsmear}. Let us also mention that the Poisson brackets (\ref{zerobrace}, \ref{PiUbrace}) are qualitatively unchanged if evaluated in an irreducible representation of the group:
\begin{equation}
\fl\qquad\eqalign{
\big\{\Pi_i[f],D^{(j,k)}\big(h_\gamma\big)\big\}=-\epsilon(f,\gamma)D^{(j,k)}\big(h_{\gamma^+}\big)\ou{D}{(j,k)}{\!\!\ast}(\tau_i)D^{(j,k)}\big(h_{\gamma^-}\big),\\
\big\{\bar\Pi_i[f],D^{(j,k)}\big(h_\gamma\big)\big\}=-\epsilon(f,\gamma)D^{(j,k)}\big(h_{\gamma^+}\big)\ou{D}{(j,k)}{\!\!\ast}(\bar{\tau}_i)D^{(j,k)}\big(h_{\gamma^-}\big).
}\end{equation}
Where $\ou{D}{(j,k)}{\!\!\ast}$ is the corresponding induced representation of the Lie algebra $\mathfrak{sl}(2,\mathbb{C})$.

Following the general strategy of loop quantum gravity the smeared versions of momentum \eref{momsmear} and connection \eref{holdef} can now naturally be turned into operators mapping all of $\mathrm{Cyl}$ onto itself. For any $\Psi\in\mathrm{Cyl}$ we define
\numparts
\begin{eqnarray}
\widehat{\Pi_i[f]}\Psi=\frac{\hbar}{\I}\big\{\Pi_i[f],\Psi\big\},\qquad
\widehat{\bar\Pi_i[f]}\Psi=\frac{\hbar}{\I}\big\{\bar\Pi_i[f],\Psi\big\},\\
\big(\ou{\big[\widehat{D^{(j,k)}(h_\gamma)}\big]}{\mu\bar\rho}{\nu\bar\sigma}\Psi\big)[A]=
\ou{\big[{D^{(j,k)}(h_\gamma[A])}\big]}{\mu\bar\rho}{\nu\bar\sigma}\Psi[A].
\end{eqnarray}
\endnumparts
If we furthermore introduce an adjointness relation in the obvious way, namely by setting for any operator $O$ on $\mathrm{Cyl}$ that
\begin{equation}
\forall \Psi,\Psi^\prime\in\mathrm{Cyl}:\langle\Psi,{O}^\ast\Psi^\prime\rangle_{\mathrm{Cyl}}:=\langle{O}\Psi,\Psi^\prime\rangle_{\mathrm{Cyl}}
\end{equation}
we observe the $\ast$-operation to do nothing but to exchange the sectors of different chirality between one another:
\numparts
\begin{eqnarray}
\widehat{\Pi_i[f]}^\ast=\widehat{\bar{\Pi}_i[f]},\\
\Big(\ou{\big[\widehat{D^{(j,k)}(h_\gamma)}\big]}{\mu\bar\rho}{\nu\bar\sigma}\Big)^\ast=\ou{\widehat{D^{(k,j)}}\big(h_\gamma\big)}{\rho\bar\mu}{\sigma\bar\nu}.
\end{eqnarray}
\endnumparts
Which induces a natural involution on $\mathrm{Cyl}$. From now on we are going to omit hats wherever possible.
%%%%%%%%%%%%%%%%%%%%%%%
%%%%%%%%%%%%%%%%%%%%%%%
%%%%%%%%%%%%%%%%%%%%%%%
\section{Calculating the Chern--Simons expectation value}\label{sect4}
\subsection{Functional integral}
After having defined the space of $SL(2,\mathbb{C})$ boundary states we are now going to investigate the vertex amplitude proposed. This is nothing but the Chern--Simons expectation value
\begin{equation}
\forall\Psi\in\mathrm{Cyl}:\mathsf{A}_{\mathrm{vertex}}[\Psi]=\int_{\mathcal{A}_{SL(2,\mathbb{C})}}\!\!\!\!\!\mathcal{D}[A]\E^{-\frac{n}{8\pi}(\beta+\I)Y[A]-\CC}\Psi[A].
\end{equation}
of the boundary spinnetwork under consideration. Applying the $SU(N)$-framework developed in \cite{FroehlichCS} to the desired case of $SL(2,\mathbb{C})$ we will straight forwardly recover all the well known results of e.g. \cite{Wittenreal, Wittenkomplex, FroehlichCS}. The only point in our calculation where we are going to significantly deviate form \cite{FroehlichCS}, and introduce something new 
concerns the integration over the non-compact gauge group, hidden in the formal integration measure $\mathcal{D}[A]$. 

For this purpose, consider a particular gauge fixing, allowing us to split \cite{bertlmann} the integration into
\qq{horizontal} and \qq{vertical} parts:
\begin{equation}
\int_{\mathcal{A}_{SL(2,\mathbb{C})}}\!\!\!\!\!\mathcal{D}[A]\Psi[A]=\int_{\mathcal{G}_{SL(2,\mathbb{C})}}\!\!\!\!\!\mathcal{D}[g]\int_{\mathcal{A}^{\mathrm{gf}}_{SL(2,\mathbb{C})}}\!\!\!\!\!\mathcal{D}[A]\Delta_{\mathrm{FP}}[A,g]\Psi[A^g].
\end{equation}
Where $\Delta_{\mathrm{FP}}$ is the Fadeev--Popov determinant associated to the change of variables, $\mathcal{G}_{SL(2,\mathbb{C})}$ denotes the \qq{fibre} generated by infinitesimal gauge transformations, and $A^g$ is the gauge transformed connection \eref{gaugetrafo}.
Insofar we restrict ourselves to gauge invariant elements
\begin{equation}
\big(U_{g^{-1}}\Psi\big)[A]:=\Psi[A^g]=\Psi[A],
\end{equation}
and choose an axial gauge fixing explicitly introduced later, the integration over $\mathcal{G}_{SL(2,\mathbb{C})}$ results in an overall constant, which can generally be absorbed into the definition of $\mathcal{D}[A]$. In loop quantum gravity we are in a more delicate situation. The reality conditions \eref{realcond} manifestly break $SL(2,\mathbb{C})$ gauge invariance.  This feature is shared with all the modern spinfoam models, where the simplicity constraints (i.e. the analog of the reality conditions) intermediately break the underlying gauge symmetry. In order to recover full gauge invariance \cite{lortzcov} one is then forced to explicitly perform the average over the group. In our case we may try to achieve this by means of the Haar measure \eref{haarmeas} just to run into the very same problems already encountered in the beginning of section \ref{sect3}. The integration over the gauge group (over each single point in $\Sigma$) results in an infinite expression, which for arbitrary spinnetwork functions cannot obviously be removed.

Once again the \emph{formal} integration measure introduced in \eref{measdef}, fulfilling a number of properties \eref{invar2} of invariance may offer a way out of this. For this purpose let $\Psi_f=f(h_{\gamma_1}[A],\dots,h_{\gamma_L}[A])$ be an element of $\mathrm{Cyl}$ containing $L$ coloured links. Define the set $\{p_1,\dots,p_N\}$ of nodes, i.e. the collection of all the final and initial points of links. By virtue of \eref{invar2} the map $P:\mathrm{Poly}_{SL(2,\mathbb{C})^L}\rightarrow\mathrm{Poly}_{SL(2,\mathbb{C})^L}$ 
\begin{equation}
\eqalign{
(Pf)(h_1,\dots,h_L)=\int_{SL(2,\mathbb{C})}\!\!\!\!\!&\mathfrak{D}(g_1)\cdots\int_{SL(2,\mathbb{C})}\!\!\!\!\!\mathfrak{D}(g_N)\\
&f\big(g_{t(1)}^{-1}h_1g_{s(1)},\dots,g_{t(L)}^{-1}h_Lg_{s(L)}\big).
}
\end{equation}
is a proper projector. Here for the $l$-th link $p_{t(l)}=\gamma_l(1)$ denotes the final (target) point, and $p_{s(l)}=\gamma_l(0)$ equals the initial point (i.e. the source).
For any $\Psi_f[A]=f(h_{\gamma_1}[A],\dots)$ the equation
\begin{equation}
P\Psi_f=\Psi_{Pf}
\end{equation}
naturally lifts $P$ to all of $\mathrm{Cyl}$, allowing us to define the formal integration over the \qq{fibre} $\mathcal{G}_{SL(2,\mathbb{C})}$:
\begin{equation}
\forall\Psi\in\mathrm{Cyl}:\int_{\mathcal{G}_{SL(2,\mathbb{C})}}\!\!\!\!\!\mathcal{D}[g]\Psi[A^g]:=\big(P\Psi\big)[A].
\end{equation}
\subsection{Axial gauge fixing}
Following the derivation of \cite{FroehlichCS} we set $\Sigma\simeq\mathbb{R}^3$ and introduce complex coordinates therein
\begin{equation}
\vec{x}=(t,x,y)\in\mathbb{R}^3\mapsto (t,z)=(t,x+\I y)\in\mathbb{R}\times\mathbb{C}.
\end{equation}
Consider the following \qq{axial} gauge condition, always possible to achieve:
\begin{equation}
A\in\mathcal{A}^{\mathrm{ax}}_{SL(2,\mathbb{C})}:A^i(\partial_{\bar{z}})=\ou{A}{i}{a}\partial^a_{\bar{z}}=0.\label{axgauge}
\end{equation}
Where we have introduced the complex valued tangent vectors (Wirtinger derivatives)
\begin{equation}
\partial_z=\frac{1}{2}\Big(\frac{\partial}{\partial x}-\I\frac{\partial}{\partial y}\Big),\quad\partial_{\bar{z}}=\frac{1}{2}\Big(\frac{\partial}{\partial x}+\I\frac{\partial}{\partial y}\Big).
\end{equation}
Notice furthermore that the axial choice corresponds only to a partial gauge fixing. Indeed any transformation generated by an analytic Lie algebra element $\partial_{\bar{z}}\Lambda^i=0$ will always preserve this condition.  
Using this particular gauge fixing the Fadeev--Popov determinant
\begin{equation}
\Delta_{\mathrm{FP}}=\mathrm{det}\frac{\delta \ou{A}{i}{a}(\vec{x})\partial^a_{\bar{z}}}{\delta \Lambda^j(\vec{x}^\prime)}\Big|_{\mathcal{A}^{\mathrm{ax}}_{SL(2,\mathbb{C})}}=\mathrm{det}\Big(\delta^i_j\partial_{\bar{z}}\delta(\vec{x}-\vec{x}^\prime)\Big)
\end{equation}
turns out to be independent \cite{bertlmann} of $g=\exp(\Lambda^i\tau_i)$ and $A$, and can thus formally be absorbed into the definition of the measure, which has to be normalised anyhow. No \qq{ghost} fields must be introduced. There is another advantage of this gauge, the cubic self interaction $A\wedge A\wedge A$ automatically vanishes, leaving us with an expression quadratic in the fields: Take the natural parametrisation
\begin{equation}
A^i=\varphi^i\di t+\pi^i\di z,
\end{equation}
of any element of $\mathcal{A}^{\mathrm{ax}}_{SL(2,\mathbb{C})}$ into complex (i.e. $\mathfrak{sl}(2,\mathbb{C})$ valued) fields $\varphi$, $\pi$, and indeed the Chern--Simons action simplifies:
\begin{equation}
Y[\varphi\di t+\pi\di z]
=-\I\int_{\mathbb{R}^3}\di^3x\big(\varphi_i\partial_{\bar{z}}\pi^i-\pi_i\partial_{\bar{z}}\varphi^i\big).\label{redactn}
\end{equation}
Where $\di^3x=\di t \di x \di y$ equals the fiducial volume element on $\mathbb{R}^3$. Introducing a formal integration measure we arrive at the gauge fixed path integral
\begin{equation}
\eqalign{
\langle\Psi\rangle_{\mathrm{ax}}:=\frac{1}{N(\lambda)}\int_{\mathcal{A}^{\mathrm{ax}}_{SL(2,\mathbb{C})}}\!\!\!\!\!
\mathcal{D}[\varphi]\mathcal{D}[\pi]&\E^{\I\lambda\int_{\mathbb{R}^3}\di^3x(\varphi_i\partial_{\bar{z}}\pi^i-\partial_{\bar{z}}\varphi_i\pi^i)-\CC}\\
&\cdot\Psi[\varphi\di t+\pi\di z].}\label{pathint}
\end{equation}
Here, there appears the complex valued coupling constant
\begin{equation}
\lambda=\frac{n}{8\pi}(\beta+\I),
\end{equation}
together with an unspecified normalisation
$N(\lambda)$. Furthermore there is the infinite dimensional integration measure formally given by
\begin{equation}
\mathcal{D}[\varphi]=\prod_{\vec{x}\in\mathbb{R}^3}\frac{\mathbbm{d}^3\varphi(\vec{x})\mathbbm{d}^3\bar{\varphi}(\vec{x})}{(2\I)^3},
\end{equation}
and equally for $\mathcal{D}[\pi]$, where $\mathbbm{d}$ denotes the functional differential. The connection to the non-gauge fixed path-integral is established by the relation
\begin{equation}
\mathsf{A}_{\mathrm{vertex}}[\Psi]=\int_{\mathcal{A}_{SL(2,\mathbb{C})}}\mathcal{D}[A]\E^{-\lambda Y[A]-\CC}\Psi[A]=\langle P\Psi\rangle_{\mathrm{ax}}.
\end{equation}
In other words, the integration over the fibre has been replaced with a projector onto the gauge invariant elements of $\mathrm{Cyl}$, and we are thus left to compute the gauge fixed integral only. We are going to achieve this, by first enquiring a finite dimensional analog of \eref{pathint} equipped with \qq{Schwinger sources} $A$ and $B$, that is the ordinary integral
\begin{equation}
\eqalign{
Z(A,B)&=\frac{\mathrm{det}(J)}{(2\pi)^{2N}}\int_{\mathbb{C}^{N}}\frac{\di^Nz\di^N\bar{z}}{(2\I)^N}\int_{\mathbb{C}^{N}}\frac{\di^N\zeta\di^N\bar{\zeta}}{(2\I)^N}\\
&\qquad\cdot\E^{\frac{\I}{2}\zeta_l\ou{G}{l}{m}z^m+A_mz^m+B^l\zeta_l-\CC}\\
&=\E^{2\I A_n\ou{(G^{-1})}{n}{m}B^m-\CC}.\label{mtvtn}
}
\end{equation}
Where $G\in GL(N,\mathbb{C})$ denotes the Jacobian
\begin{equation}
J=\pmatrix{ \mathfrak{Re}(G) & -\mathfrak{Im}(G) \cr \mathfrak{Im}(G) & \mathfrak{Re}(G)}
\end{equation}
appearing in the normalisation of the \qq{partition} function $Z(A,B)$. In the infinite dimensional case $G$ is explicitly given as an integration kernel:
\numparts
\begin{equation}
\ou{G}{j}{j^\prime}(\vec{x},\vec{x}^\prime)=4\lambda\delta^j_{j^\prime}\partial_{\bar{z}}\delta^3(\vec{x}-\vec{x}^\prime).
\end{equation}
Choosing appropriate falloff conditions the inverse (i.e. the Green function) is determined to be
\begin{equation}
\ou{\big[G^{-1}\big]}{j}{j^\prime}(\vec{x},\vec{x}^\prime)=\frac{1}{4\pi\lambda}\delta^j_{j^\prime}\delta(t-t^\prime)\frac{1}{z-z^\prime}
\end{equation}
\endnumparts
Which follows from the distributional equation $4\partial_z\partial_{\bar{z}}\mathrm{ln}|z|=2\pi\delta(z)$.
In order to calculate arbitrary $n$-point functions it proves useful to study the finite dimensional analog of \eref{mtvtn}. The action is quadratic in the \qq{fields} and all the $n$-point functions are already determined by the two-point functions, itself being proportional to the \qq{Green function}  $G^{-1}$. Lifting the results from the finite dimensional case to the infinite dimensional we are able to introduce the correlation functions: 
\begin{equation}
\eqalign{
\big\langle\pi^j(\vec{x})\varphi_{j^\prime}(\vec{x}^\prime)\big\rangle_{\mathrm{ax}}=
\frac{\I}{2\pi\lambda}\delta^j_{j^\prime}\delta(t-t^\prime)\frac{1}{z-z^\prime},\\
\big\langle\bar\pi^j(\vec{x})\bar\varphi_{j^\prime}(\vec{x}^\prime)\big\rangle_{\mathrm{ax}}=
\frac{\I}{2\pi\bar\lambda}\delta^j_{j^\prime}\delta(t-t^\prime)\frac{1}{\bar{z}-\bar{z}^\prime}.
}\label{corrfunc}
\end{equation}
All the other two point functions vanish. 
\subsection{Knizhnik--Zamolodchikov equations}
For the present let us think about the following reduced problem. Consider a graph $\Gamma$ containing a number of $L$ \qq{vertical} lines nowhere intersecting. These are links allowing for parametrisation in \qq{time} $t$, that is:
\begin{equation}
\gamma_I:t\mapsto \big(t,z_I(t)\big),\quad\mbox{and:}\quad
\forall t,I\neq J:z_I(t)\neq z_J(t)\label{braidspar}
\end{equation}
Evaluated at connections which fulfil the axial gauge condition \eref{axgauge} the holonomy turns into
\begin{equation}
h_{\gamma_I(t)}=\boldsymbol{P}\exp\Big(-\int_0^t\di t^\prime\big[\varphi^i\big(t^\prime,z_I(t^\prime)\big)+\pi^i\big(t^\prime,z_I(t^\prime)\big)\dot{z}_I(t^\prime)\big]\tau_i \Big).
\end{equation}
To further simplify our calculations let us restrict all the holonomies to be coloured with the fundamental---that is the $(1/2,0)$-representation (we'll comment on more general cases later):
\begin{equation}
\ou{[H_{\Gamma(t)}]}{\alpha_1\dots\alpha_L}{\beta_1\dots\beta_L}=\big\langle \ou{[h_{\gamma_{1}(t)}]}{\alpha_1}{\beta_1}\dots\ou{[h_{\gamma_{L}(t)}]}{\alpha_L}{\beta_L}\big\rangle_{\mathrm{ax}}.
\end{equation}
Following \cite{FroehlichCS}, our strategy concerns the derivation of a set of first order differential equations for $H_{\Gamma(t)}$, later allowing us to deduce the full functional integral. These are the Knizhnik--Zamolodchikov equations, extensively studied in e.g. \cite{CFT}. Consider first Wick's theorem, which is nothing but
\numparts
\begin{eqnarray}
\big\langle\varphi_i(t,z_I(t))\Psi\big\rangle_{\mathrm{ax}} & = &  \frac{1}{4\pi\lambda}\int_{\mathbb{C}}\frac{\di z\wedge\di\bar{z}}{z-z_I(t)+\varepsilon}\Big\langle\frac{\delta}{\delta\pi^i(t,z)}\Psi\Big\rangle_{\mathrm{ax}}\label{wick1}\\
\big\langle\pi^i(t,z_I(t))\Psi\big\rangle_{\mathrm{ax}} & = &  \frac{1}{4\pi\lambda}\int_{\mathbb{C}}\frac{\di z\wedge\di\bar{z}}{z_I(t)-z-\varepsilon}\Big\langle\frac{\delta}{\delta\varphi_i(t,z)}\Psi\Big\rangle_{\mathrm{ax}}\label{wick2}
\end{eqnarray}
\endnumparts
Where we have introduced an $\varepsilon\rightarrow 0$ description intended to cancel infinities otherwise appearing. If we calculate the derivative in time we find:
\begin{equation}
\eqalign{
\frac{\di}{\di t}H_{\Gamma(t)} = - \sum_{I=1}^L\Big\langle h_{{\gamma_1}(t)}\otimes\dots\\
\quad\dots\otimes\big(\varphi^i(t,z_I(t))+\pi^i(t,z_I(t))\dot{z}_I(t)\big)\tau_ih_{\gamma_I(t)}\otimes\dots\otimes h_{\gamma_L(t)}\Big\rangle_{\mathrm{ax}}.}\label{timeder}
\end{equation}
By Wick's theorem the multiplication by $\varphi$ turns into a functional derivative related to nothing but the flux through the complex plane at time $t$:
\begin{equation}
\fl\quad\eqalign{\Big\langle& \varphi_i(t,z_I(t))\big(h_{{\gamma_1}(t)}\otimes\dots\otimes h_{\gamma_L(t)}\big)\Big\rangle_{\mathrm{ax}}  = \\
&=  \frac{1}{4\pi\lambda}\int_{\mathbb{C}}\frac{\di z\wedge\di\bar{z}}{z-z_I(t)+\varepsilon}\Big\langle\frac{\delta}{\delta\pi^i(t,z)}\big(h_{{\gamma_1}(t)}\otimes\dots\otimes h_{\gamma_L(t)}\big)\Big\rangle_{\mathrm{ax}}=\\
&=\frac{1}{4\pi\I\lambda}\sum_{J=1}^L\frac{\dot{z}_J(t)}{z_J(t)-z_I(t)+\varepsilon}
\big\langle h_{\gamma_1(t)}\otimes\dots\otimes\tau_ih_{\gamma_J(t)}\otimes\dots\otimes h_{\gamma_L(t)}\big\rangle_{\mathrm{ax}}}
\end{equation}
Here there appears an extra factor of $1/2$, which can be traced back to the regularisation of the delta function $\int_0^t\di t^\prime\delta(t-t^\prime)=\frac{1}{2}$. For the multiplication by $\pi$ the calculation is essentially the same:
\begin{equation}
\fl\quad\eqalign{\Big\langle& \pi^i(t,z_I(t))\big(h_{{\gamma_1}(t)}\otimes\dots\otimes h_{\gamma_L(t)}\big)\Big\rangle_{\mathrm{ax}}  =  \\
&=\frac{1}{4\pi\lambda}\int_{\mathbb{C}}\frac{\di z\wedge\di\bar{z}}{z_I(t)-z-\varepsilon}\Big\langle\frac{\delta}{\delta\varphi_i(t,z)}\big(h_{{\gamma_1}(t)}\otimes\dots\otimes h_{\gamma_L(t)}\big)\Big\rangle_{\mathrm{ax}}=\\
&=\frac{1}{4\pi\I\lambda}\sum_{J=1}^L\frac{1}{z_I(t)-z_J(t)-\varepsilon}
\big\langle h_{\gamma_1(t)}\otimes\dots\otimes\tau_ih_{\gamma_J(t)}\otimes\dots\otimes h_{\gamma_L(t)}\big\rangle_{\mathrm{ax}}}
\end{equation}
Inserting both results into \eref{timeder} the defining differential equations turns into
\begin{equation}
\fl\qquad\eqalign{
\frac{\di}{\di t}H_{\Gamma(t)} = -\frac{1}{4\pi\I\lambda}&\sum_{I<J}\bigg[\frac{\dot{z}_I(t)-\dot{z}_J(t)}{z_I(t)-z_J(t)+\varepsilon}+\frac{\dot{z}_I(t)-\dot{z}_J(t)}{z_I(t)-z_J(t)-\varepsilon}\bigg]\cdot\\
&\cdot\Big\langle h_{\gamma_1(t)}
\dots\otimes\tau_i h_{\gamma_I(t)}\otimes\dots
\tau^i h_{\gamma_J(t)}\otimes\dots\otimes h_{\gamma_L(t)}\Big\rangle_{\mathrm{ax}}}
\end{equation}
Where the otherwise ill-defined expression $\frac{\dot{z}_I-\dot{z}_I}{z_I-z_I}$ cancels by virtue of the $\varepsilon$-description. Introducing for $I<J$ the grasping, i.e. the ${2^L}\times{2^L}$ dimensional $\mathbb{C}$-valued matrix:
\begin{equation}
\Omega_{IJ}\equiv\Omega_{JI}=\mathds{1}\otimes\dots\otimes\underset{I\mathrm{-th}}{\tau_i}\otimes\dots\otimes\underset{J\mathrm{-th}}{\tau^i}\otimes\dots\otimes\mathds{1}
\end{equation}
and performing the limit of $\varepsilon\rightarrow 0$ the differential equation for $H_\Gamma$ turns into:
\begin{equation}
\frac{\di}{\di t}H_{\Gamma(t)}=-\frac{1}{2\pi\I\lambda}\sum_{I<J}\frac{\dot{z}_I(t)-\dot{z}_J(t)}{z_I(t)-z_J(t)}\Omega_{IJ}H_{\Gamma(t)}\label{exhol}
\end{equation}
Demanding the normalisation $\langle 1\rangle_{\mathrm{ax}}=1$, thereby implicitly fixing the measure $\mathcal{D}[A]$, the initial condition turns into the following:
\begin{equation}
H_{\Gamma(0)}=\underbrace{\mathds{1}\otimes\dots\otimes\mathds{1}}_{L-\mathrm{times}}
\end{equation}
The grasping matrices $\Omega_{IJ}$ for indices $I,J,M,N$ being all different obey the infinitesimal skein relations
\begin{equation}
[\Omega_{IJ},\Omega_{MN}]=[\Omega_{IJ},\Omega_{IM}+\Omega_{MJ}]=0.\label{infskein}
\end{equation}
which for the anti-selfdual sector (generated by $\bar{\Omega}_{12}=\bar{\tau}_i\otimes\bar{\tau}^i\otimes\bar{\mathds{1}}\otimes\dots$ and so on) are equally satisfied.
In fact the derivations performed can immediately be translated to the sector of opposite chirality. %We are intersted to compute:
%\begin{equation}
%H_{\Gamma(t)}=\big\langle h_{\gamma_L(t)}\otimes\dots\otimes h_{\gamma_L(t)}\otimes \bar{h}_{\gamma_{L+1}(t)}\otimes\dots\otimes \bar{h}_{\gamma_{L+L^\prime}(t)}\big\rangle_{\mathrm{ax}}
%\end{equation}
Again the two sectors decouple.  If ${\gamma_{L+1}},\dots,{\gamma_{L+L^\prime}}$ are additional edges coloured with the complex conjugate i.e. $(0,1/2)$ representation, 
the defining differential equation becomes:
\begin{equation}
\eqalign{
\frac{\di}{\di t}H_{\Gamma(t)}=&-\frac{1}{2\pi\I\lambda}\sum_{1\leq I<J\leq L}\frac{\dot{z}_I(t)-\dot{z}_J(t)}{z_I(t)-z_J(t)}\Omega_{IJ}H_{\Gamma(t)}+\\
&-\frac{1}{2\pi\I\bar\lambda}\sum_{L<R<S\leq L+L^\prime}\frac{\dot{\bar{z}}_R(t)-\dot{\bar{z}}_S(t)}{\bar{z}_R(t)-\bar{z}_S(t)}\bar\Omega_{RS}H_{\Gamma(t)}\\
\mbox{and:}\;H_{\Gamma(0)}=&\underbrace{\mathds{1}\otimes\dots\otimes\mathds{1}}_{L-\mathrm{times}}\otimes\underbrace{\bar\mathds{1}\otimes\dots\otimes\bar\mathds{1}}_{L^\prime-\mathrm{times}}
\label{exhol2}
}
\end{equation}
Equation \eref{exhol2} allows for a  vivid physical interpretation \cite{FroehlichCS}. Consider the punctured space 
\begin{equation}
M_{L+L^\prime}:=\{(z_1,\dots,z_{L+L^\prime})\in\mathbb{C}^{L+L^\prime}|\forall I\neq J:z_I\neq z_J\},
\end{equation}
equipped with the following (matrix valued) covariant derivative:
\begin{equation}
\eqalign{
\nabla^{I}=\frac{\partial}{\partial z_I}+\frac{1}{2\pi\I\lambda}\sum_{J=1}^{L}\Big|_{J\neq I}\frac{1}{z_I-z_J}\Omega_{IJ}\label{covdirav}\\
\bar{\nabla}^{R}=\frac{\partial}{\partial \bar{z}_R}+\frac{1}{2\pi\I\bar{\lambda}}\sum_{S={L+1}}^{L+L^\prime}\Big|_{S\neq R}\frac{1}{\bar{z}_R-\bar{z}_S}\bar{\Omega}_{RS}}
\end{equation}
By virtue of the infinitesimal skein relation \eref{infskein} (and the corresponding equations for the complex conjugate sector) the curvature vanishes:
\begin{equation}
\eqalign{
\forall I,J\in\{1,\dots,L\}; R,S\in\{L+1,\dots,L+L^\prime\}\\
\big[\nabla^I,\nabla^J\big]=\big[\bar\nabla^R,\nabla^J\big]=\big[\bar\nabla^R,\bar\nabla^S\big]=0\label{zerocurv}
}
\end{equation}
And most importantly \cite{FroehlichCS, Wittenkomplex}, the defining differential equation \eref{exhol2} is nothing but the parallel transport along the path $\Gamma(t)=(z_1(t),\dots,z_{L+L^\prime}(t))$ in $M_{L+L^\prime}$:
\begin{equation}
\sum_{I=1}^L\dot{z}_I(t)\nabla^IH_{\Gamma(t)}+\sum_{R=L+1}^{L+L^\prime}\dot{\bar{z}}_R(t)\bar\nabla^RH_{\Gamma(t)}=0
\end{equation}
But the connection is flat implying that the parallel transport cannot depend if $\Gamma$ is carefully deformed. Indeed $H_\Gamma$ is a proper functional of the whole homotopy class $[\Gamma]$ of $\Gamma$, inducing a matrix representation of the corresponding homotopy group $\pi_1(M_{L+L^\prime})$ of the punctured space. This is nothing but Artin's braid group \cite{Lerda} of $L+L^\prime$ strands. 
There is an important subtlety here. The Chern--Simons functional is 
invariant $Y[\varphi^\ast A]=Y[A]$ under orientation preserving diffeomorphisms $\varphi:\Sigma\rightarrow\Sigma$. For any spin-network function $\Psi_\Gamma$ we naively expect this to happen for the expectation value too, that is
\begin{equation}
\int_{\mathcal{A}_{SL(2,\mathbb{C})}}\!\!\!\!\!\mathcal{D}[A]\overline{\Omega[A]}\Psi_\Gamma[A]\stackrel{?}{=}
\int_{\mathcal{A}_{SL(2,\mathbb{C})}}\!\!\!\!\!\mathcal{D}[A]\overline{\Omega[A]}\Psi_{\varphi(\Gamma)}[A],
\end{equation}
which is however not true. In fact it is the measure $\mathcal{D}[A]$ which by means of the decomposition $\mathbb{R}^3\simeq\mathbb{R}\times\mathbb{C}$ into \qq{time} and \qq{space} implicitly breaks diffeomorphism invariance. Luckily, this violation of diffeomorphism invariance is not very harmful, it may occur only if the diffeomorphism considered moves the turning points of the spinnetwork function $\Psi_\Gamma$. A turning point appears on a given link $s\mapsto(t(s),z(s))$ wherever the map $s\mapsto t(s)$ has a local maximum (minimum). Notice furthermore that in the parametrisation chosen \eref{braidspar} any turning point happens to be an intersection of two links $I\neq J$ at their common boundaries, e.g. $z_I(0)=z_J(0)$. On these points the differential equation \eref{exhol2} becomes singular, making a regularisation scheme necessary, implicitly sketched during the next section. 

The vanishing of the curvature implies also the solvability of \eref{exhol2} by a system of $L+L^\prime$ partial differential equations. These turn out to be the Knizhnik--Zamolodchikov equations for both the holomorphic and anti-holomorphic blocks:
\begin{equation}
\eqalign{
\nabla^I\Psi(z_1,\dots,z_L)=0,&\quad\bar\nabla^R\Xi(\bar{z}_{L+1},\dots,\bar{z}_{L+L^\prime})=0,\\
\Phi\big(z_1(0),\dots,z_L(0)\big)=\phi,&\quad\Xi\big(\bar{z}_{L+1}(0),\dots,\bar{z}_{L+L^\prime}(0)\big)=\xi.
}
\end{equation}
Where $\Psi$, and $\Xi$ take values in $\bigotimes^L(1/2,0)$ and $\bigotimes^{L^\prime}(0,1/2)$ respectively. Given solutions (e.g. for the four point \cite{CFT, FroehlichCS} correlation function) we could immediately construct the Chern--Simons expectation value associated:
\begin{equation}
\eqalign{
H_{\Gamma(t)}\phi\otimes\xi&=\Phi\big(z_1(t),\dots,z_L(t)\big)\otimes\Xi\big(\bar{z}_{L+1}(t),\dots,\bar{z}_{L+L^\prime}(t)\big)\\
&=\big\langle h_{\gamma_L(t)}\otimes\dots\otimes h_{\gamma_L(t)}\otimes\\
&\qquad\otimes\bar{h}_{\gamma_{L+1}(t)}\otimes\dots\otimes \bar{h}_{\gamma_{L+L^\prime}(t)}\big\rangle_{\mathrm{ax}}\cdot\phi\otimes\xi
}\label{prodprop}\end{equation}
\subsection{Skein relations}
In the last section we saw that the derivation of the Chern--Simons expectation value reduces to the calculation of the parallel transport of a certain matrix valued connection. This connection is flat \eref{zerocurv} implying that the Chern--Simons expectation value is invariant if the links are carefully deformed (e.g. without crossing one another). 

It was Witten who first proved in \cite{Wittenreal} that the Chern--Simons expectation value turns out to be fully determined by a number of skein relations. These equations which we will introduce in a moment immediately relate the Chern--Simons expectation value of a product of non intersecting Wilson lines to the Jones polynomial of knot theory \cite{kauffmann}. 

Let us start with a rather trivial case, consider the path $\Gamma^0:[0,1]\rightarrow M_2$ formed by two parallel lines, parametrised according to
\begin{equation}
\Gamma^0:t\mapsto\big(z_1^0(t),z_2^0(t)\big)=(-1,1).
\end{equation}
The corresponding holonomy happens to be the identity
\begin{equation}
H_{\Gamma^0(t)}=\mathds{1}\otimes\mathds{1}.
\end{equation}
In order to arrive at the skein relations determining the Chern--Simons expectation value, we need to study more sophisticated graphs. Consider two links encircling one another in (counter-)clockwise orientation:
\begin{equation}
\eqalign{
\Gamma^+(t):\big(z_1^+(t),z_2^+(t)\big)=\big(-\E^{+\I\pi t},\E^{+\I\pi t}\big),\\
\Gamma^-(t):\big(z_1^-(t),z_2^-(t)\big)=\big(-\E^{-\I\pi t},\E^{-\I\pi t}\big).
}
\end{equation}
If we set $z^\pm(t)=z_1^\pm(t)-z_2^\pm(t)$ together with $\Omega=\tau_i\otimes\tau^i$, we obtain the defining differential equation for the holonomies:
\begin{equation}
\frac{\di}{\di t}H_{\Gamma^\pm(t)}=\mp\frac{1}{2\lambda}\Omega H_{\Gamma^\pm(t)}.
\end{equation}
The initial condition $H_{\Gamma^\pm(0)}=\mathds{1}\otimes\mathds{1}$ determines the solution to be:
\begin{equation}
H_{\Gamma^\pm(t)}=\exp\Big(\mp\frac{t}{2\lambda}\Omega\Big).
\end{equation}
Define the transposition operator
\begin{equation}
X=\frac{1}{2}\mathds{1}\otimes\mathds{1}-2\Omega,
\end{equation}
which acts onto $\mathbb{C}^2\otimes\mathbb{C}^2$ according to
\begin{equation}
\forall\varphi_1,\varphi_2\in\mathbb{C}^2:X\varphi_1\otimes\varphi_2=\varphi_2\otimes\varphi_1.
\end{equation}
We find
\begin{equation}
\eqalign{
H_{\Gamma^\pm(t)}&=\E^{\mp\frac{t}{8\lambda}}\E^{\pm\frac{t}{4\lambda}X}=\\
&=\E^{\mp\frac{t}{8\lambda}}\mathrm{ch}\Big(\frac{t}{4\lambda}\Big)\mathds{1}\otimes\mathds{1}\pm\E^{\mp\frac{t}{8\lambda}}\mathrm{sh}\Big(\frac{t}{4\lambda}\Big)X.}
\end{equation}
This in turn implies the skein relation
\begin{equation}
\E^{\frac{1}{8\lambda}}H_{\Gamma^+(1)}-\E^{-\frac{1}{8\lambda}}H_{\Gamma^-(1)}=2\mathrm{sh}\Big(\frac{1}{4\lambda}\Big)X.
\end{equation}
If we recover spinorial indices we find this equation to be
\begin{equation}
\E^{\frac{1}{8\lambda}}\ou{\big[H_{\Gamma^+(1)}\big]}{\alpha\beta}{\mu\nu}-\E^{-\frac{1}{8\lambda}}\ou{\big[H_{\Gamma^-(1)}]}{\alpha\beta}{\mu\nu}=2\mathrm{sh}\Big(\frac{1}{4\lambda}\Big)\delta^\alpha_\nu\delta^\beta_\mu.
\end{equation}
Introducing a natural graphical notation these relations become increasingly more transparent. Let us define:
\numparts
\begin{eqnarray}
\big\langle{{}^\beta_\mu\LZ^\alpha_\nu}\big\rangle_{\mathrm{ax}} & := & \ou{\big[H_{\Gamma^+(1)}\big]}{\alpha\beta}{\mu\nu}=\big\langle\ou{[h_{\gamma^+_1(1)}]}{\alpha}{\mu}\ou{[h_{\gamma^+_2(1)}]}{\beta}{\nu}\big\rangle_{\mathrm{ax}},\label{exa}\\
\big\langle{{}^\beta_\mu\LX^\alpha_\nu}\big\rangle_{\mathrm{ax}} & := & \ou{\big[H_{\Gamma^-(1)}\big]}{\alpha\beta}{\mu\nu}=\big\langle\ou{[h_{\gamma^-_1(1)}]}{\alpha}{\mu}\ou{[h_{\gamma^-_2(1)}]}{\beta}{\nu}\big\rangle_{\mathrm{ax}},\label{exb}\\
\big\langle{{}^\alpha_\mu\Lii^\beta_\nu}\big\rangle_{\mathrm{ax}} & := & \ou{\big[H_{\Gamma^0(1)}\big]}{\alpha\beta}{\mu\nu}=\big\langle\ou{[h_{\gamma^0_1(1)}]}{\alpha}{\mu}\ou{[h_{\gamma^0_2(1)}]}{\beta}{\nu}\big\rangle_{\mathrm{ax}}=\delta^\alpha_\mu\delta^\beta_\nu.\label{exc}
\end{eqnarray}
\endnumparts
Notice the index-structure in \eref{exb} and \eref{exc}. Consider the deformation parameter
\begin{equation}
\boxed{A=\E^{\frac{1}{8\lambda}}=\E^{\frac{\pi}{n}\frac{1}{\beta+\I}},}\label{defpar}
\end{equation}
allowing us to compactly write the skein relations according to
\begin{equation}
A\big\langle{{}^\beta_\mu\LZ^\alpha_\nu}\big\rangle_{\mathrm{ax}}-A^{-1}\big\langle{{{}^\beta_\mu\LX^\alpha_\nu}}\big\rangle_{\mathrm{ax}}=(A^2-A^{-2})\big\langle{{{}^\beta_\mu\Lii^\alpha_\nu}}\big\rangle_{\mathrm{ax}}.
\end{equation}
We are now ready to drop the decorating indices in order to find
\begin{equation}
A\big\langle{\LZ}\big\rangle_{\mathrm{ax}}-A^{-1}\big\langle{{\LX}}\big\rangle_{\mathrm{ax}}=(A^2-A^{-2})\big\langle{{\Lii}}\big\rangle_{\mathrm{ax}}.\label{skein1}
\end{equation}
Let us furthermore assume that the functional is invariant under \qq{rigid} rotations, which are symmetries of the fiducial background metric on $\mathbb{R}^3$, e.g.:
\begin{equation}
\big\langle{\Lachtv}\big\rangle_{\mathrm{ax}}=\big\langle{\Lachth}\big\rangle_{\mathrm{ax}}.
\end{equation}
By this assumption, after having performed a \qq{blackboard rotation} by an angle of 90° we obtain from \eref{skein1} that:
\begin{equation}
A\big\langle{\LX}\big\rangle_{\mathrm{ax}}-A^{-1}\big\langle{\LZ}\big\rangle_{\mathrm{ax}}=(A^2-A^{-2})\big\langle{\Lun}\big\rangle_{\mathrm{ax}}.\label{skein2}
\end{equation}
This is in fact a highly non trivial step. On the right hand side there appear two extrema (a maximum and a minimum of the two respective links) where the defining differential equation \eref{exhol2} of the holonomy becomes singular. But the left hand side of \eref{skein2} is perfectly regular, and can thus naturally be viewed as a proper regularisation of the right.

An obvious algebraic manipulation brings these relations into the increasingly more useful form of
\begin{equation}\boxed{\eqalign{
\big\langle{\LX}\big\rangle_{\mathrm{ax}}=A^{-1}\big\langle{\Lii}\big\rangle_{\mathrm{ax}}+A\big\langle{\Lun}\big\rangle_{\mathrm{ax}},\\
\big\langle{\LZ}\big\rangle_{\mathrm{ax}}=A^{-1}\big\langle{\Lun}\big\rangle_{\mathrm{ax}}+A\big\langle{\Lii}\big\rangle_{\mathrm{ax}}.\label{cross1}
}}\end{equation}
As a motivating example, let us perform a typical calculation; consider the expectation value for one single Wilson loop. The curvature of the matrix valued connection \eref{covdirav} vanishes, therefore we first find that:
\begin{equation}
\big\langle{\Linf}\big\rangle_{\mathrm{ax}}=\big\langle{\Lo}\big\rangle_{\mathrm{ax}}\label{nomcop}
\end{equation}
If we fix the overall normalisation: 
\begin{equation}
\big\langle 1\big\rangle_{\mathrm{ax}}=1,\label{norm}
\end{equation}
and repeatedly use the skein relations \eref{cross1}, together with $\langle\Lo\,\Lo\rangle_{\mathrm{ax}}=\langle\Lo\rangle_{\mathrm{ax}}\langle\Lo\rangle_{\mathrm{ax}}$ (itself being motivated by a kind of \qq{surgery} \cite{Wittenreal} of the path integral) we find that:
\begin{equation}
\eqalign{
\big\langle{\Linf}\big\rangle_{\mathrm{ax}}&=A^{-1}\big\langle{\Lachtv}\big\rangle_{\mathrm{ax}}+A\big\langle{\Lo}\big\rangle_{\mathrm{ax}}\big\langle{\Lachtv}\big\rangle_{\mathrm{ax}}\\
&=A^{-2}\langle\Lo\rangle_{\mathrm{ax}}^2+\langle\Lo\rangle_{\mathrm{ax}}+\langle\Lo\rangle_{\mathrm{ax}}^3+A^2\langle\Lo\rangle_{\mathrm{ax}}^2
}
\end{equation}
Comparison with \eref{nomcop} fixes the normalisation of the unknot to be:
\begin{equation}
\boxed{\big\langle{\Lo}\big\rangle_{\mathrm{ax}}=-A^2-A^{-2}.}
\end{equation}
Using an appropriate regularisation procedure this result can also be derived \cite{FroehlichCS} directly from the 4-point Knizhnik--Zamolodchikov \cite{CFT} equations.
Notice also that
\begin{equation}
\big\langle{}^\alpha_\beta\knot\big\rangle_{\mathrm{ax}}=-A^{-3}\big\langle{}_\beta\Li{}^\alpha\big\rangle_{\mathrm{ax}}\equiv-A^{-3}\delta^\alpha_\beta.
\end{equation} 
Which explicitly shows that $\langle\,\cdot\,\rangle_{\mathrm{ax}}$ fails to be a true topological invariant. In fact, in order to determine the Chern--Simons expectation value of any product of Wilson loops a certain framing---that is a particular embedding of the collection of loops into $\mathbb{R}\times\mathbb{C}$---needs to be fixed once and for all.

Next, we should comment on the sector of opposite chirality. The skein relations turn out to be qualitatively unchanged, just the deformation parameter is replaced by its inverse complex conjugate:
\begin{equation}\boxed{\eqalign{
\big\langle{\RX}\big\rangle_{\mathrm{ax}}=\bar{A}\big\langle{\Rii}\big\rangle_{\mathrm{ax}}+\bar{A}^{-1}\big\langle{\Run}\big\rangle_{\mathrm{ax}},\\
\big\langle{\RZ}\big\rangle_{\mathrm{ax}}=\bar{A}\big\langle{\Run}\big\rangle_{\mathrm{ax}}+\bar{A}^{-1}\big\langle{\Rii}\big\rangle_{\mathrm{ax}}.\label{cross2}
}}\end{equation}
Where we have chosen another colour in order to distinguish the right-handed sector from the left, e.g.
\begin{equation}
\big\langle{{}^{\bar\alpha}_{\bar\mu}\Rii^{\bar\beta}_{\bar\nu}}\big\rangle_{\mathrm{ax}} =\ou{\big[H_{\Gamma^0(1)}\big]}{\bar\alpha\bar\beta}{\bar\mu\bar\nu}=\big\langle\ou{[\bar{h}_{\gamma^0_1(1)}]}{\bar\alpha}{\bar\mu}\ou{[\bar{h}_{\gamma^0_2(1)}]}{\bar\beta}{\bar\nu}\big\rangle_{\mathrm{ax}}=\delta^{\bar\alpha}_{\bar\mu}\delta^{\bar\beta}_{\bar\nu}.
\end{equation}
The expectation value of one single Wilson loop turns out to be
\begin{equation}
\boxed{\big\langle \Ro\big\rangle_{\mathrm{ax}}=-\bar{A}^2-\bar{A}^{-2}.}
\end{equation}
Furthermore, from equation \eref{prodprop} implying that the different sectors completely factorise, we can immediately conclude that any crossing between links of opposite chirality can trivially be moved, that is
\begin{equation}
\boxed{\big\langle\mixi\big\rangle_{\mathrm{ax}}=\big\langle\mixii\big\rangle_{\mathrm{ax}}.}
\end{equation}
This in turn finishes the collection of skein relations needed to calculate the Chern--Simons expectation value of any product of framed, non-intersecting Wilson lines.
\section{Outlook towards subsequent work}\label{sect5}
This section gives a comprehensive outlook of what needs to be done next. Leaving most of the mathematical details to a subsequent paper we focus on the general ideas.
 \subsection{Relation to $q$-deformed $SL(2,\mathbb{C})$ group and the case of intersecting links}
For the case of $SU(2)$ the corresponding Chern--Simons expectation value is closely related to a certain quantum deformation \cite{GuadagniniCS} of the group. Something similar is possible for $SL(2,\mathbb{C})$. In fact, let us first study a deformation of $SL(2,\mathbb{C})$ preserving the chiral structure of the group. This is done in close analogy to the deformation of the $SU(2)$ group. Consider the
ladder operators
\begin{equation} 
T_{\pm}=T_1\pm\I T_2,\quad\bar{T}_{\pm}=\bar{T}_1\pm\I\bar{T}_2
\end{equation}
for both the sectors of left and right chirality.
The commutation relations of the complexified Lorentz-algebra (\ref{comm1}, \ref{comm2}, \ref{comm3}) turn into:
\numparts
\begin{eqnarray}
\big[T_3,T_\pm\big]=\pm T_{\pm}, &\quad& \big[\bar{T}_3,\bar{T}_\pm\big]=\pm \bar{T}_{\pm},\label{comm4}\\
\big[T_+,T_-\big]=2T_3, &\quad& \big[\bar{T}_+,\bar{T}_-\big]=2\bar{T}_3\label{comm5}.
\end{eqnarray}
\endnumparts 
Exchanging the sectors of opposite chirality 
\begin{equation}
(T_\pm)^\ast=\bar{T}_\mp,\quad
(T_3)^\ast=\bar{T}_3\label{adrel}
\end{equation}
we naturally introduce an involution on the free algebra generated by finite complex linear cobinations of products of $T_\pm,T_3$, $\bar{T}_\pm$, $\bar{T}_3$ modulo the commutation relations (\ref{comm4}, \ref{comm5}).
Where for elements $A,B$ of this algebra and $\lambda\in\mathbb{C}$ we have that $(AB)^\ast=B^\ast A^\ast$, $(A^\ast)^\ast=A$ and $(\lambda A+B)^\ast=\bar\lambda A^\ast+B^\ast$. This algebra allows for immediate deformation in the obvious way:
\numparts
\begin{eqnarray}
\big[T_3,T_\pm\big]=\pm T_{\pm}, &\quad& \big[\bar{T}_3,\bar{T}_\pm\big]=\pm \bar{T}_{\pm},\\
\big[T_+,T_-\big]=\frac{\E^{2zT_3}-\E^{-2zT_3}}{\E^z-\E^{-z}}, &\quad& \big[\bar{T}_+,\bar{T}_-\big]=\frac{\E^{2\bar{z}\bar{T}_3}-\E^{-2\bar{z}\bar{T}_3}}{\E^{\bar{z}}-\E^{-\bar{z}}}.
\end{eqnarray}
\endnumparts
and the adjointness relations \eref{adrel} remain unchanged, still consistently exchanging the sectors of opposite chirality. The corresponding free algebra gives a natural quantum deformation of the Lorentz group.
And
\begin{equation}
A=\E^{\frac{z}{2}}\equiv\sqrt{q},\quad z=\frac{2\pi}{n}\frac{1}{\beta+\I}\in\mathbb{C}
\end{equation}
defines the relation to the deformation parameter found within the previous section. Notice that the deformation parameter is generally complex, something which is impossible to achieve for the case of $SU(2)$.

It is not very hard to show---a task which we will leave to a subsequent paper though---that the skein relations obtained from the Chern--Simons theory (\ref{cross1}, \ref{cross2}) can naturally be recovered from this \qq{chiral} deformation of the Lorentz group. 

One may now ask why this (still claimed) relation to a certain \qq{quantum group} may have physical relevance. In our opinion the main answer to this question is this: We are interested in calculating the Chern--Simons expectation value for intersecting loops. The defining differential equations \eref{exhol2} becomes however singular on the intersection points (these are the nodes of the spinnetwork function), making a regularisation necessary. This regularisation may naturally be achieved by quantum groups. In fact, this was already implicitly done in \cite{qdef1, qdef2}. All what one may need to do is to replace the spin network function considered by its corresponding quantum deformation, where on the intersection points the $SL(2,\mathbb{C})$ Wenzel--Jones projectors (generally used to construct the intertwining tensors on the nodes) for both sectors of opposite chirality are replaced by their respective quantum deformations. Which would immediately allow us to study spinnetwork functions of arbitrary colouring. The single ambiguity left would then concern the usual framing dependence, a general feature of Chern--Simons theory well known from its very beginning \cite{Wittenreal}.

\subsection{General strategy towards the reality conditions}
The reality conditions \eref{realcond} remain the main stumbling block preventing a clean definition of the model. Let us illustrate a possible way to solve them, in the context of a reduced example. Consider the smeared versions of flux $E_i[f]$ and reality conditions $C_i[f]$, as defined in (\ref{Edef}, \ref{realcond}) along a single surface $f$ in the sense of \eref{momsmear}. Construct the following scalar constraint:
\begin{equation}
\eqalign{
\boldsymbol{C}[f]&=C_i[f]E^i[f]=\\
&=\I\frac{\ellp^4}{\hbar^2}\Big(\frac{\beta^2}{(\beta+\I)^2}\Pi_i[f]\Pi^i[f]-\frac{\beta^2}{(\beta-\I)^2}\bar\Pi_i[f]\bar\Pi^i[f]\Big)\stackrel{!}{=}0\label{scalcons}
}
\end{equation}
Notice that the right hand side of this equation is manifestly $SL(2,\mathbb{C})$ invariant.

Let us turn towards the corresponding quantum constraint, for this purpose consider one single link coloured by the $(j_\ell,j_r)$ representation space. In quantum theory, by replacing the squared fluxes $\Pi[f]^2$ and $\bar\Pi[f]^2$ by the respective eigenvalues $C(j_{\ell,r})=j_{\ell,r}(j_{\ell,r}+1)$ of the two Casimir operators of $SL(2,\mathbb{C})$ equation \eref{scalcons} turns into:
\begin{equation}
(\beta^2-1)\Big(C(j_\ell)-C(j_r)\Big)
-2\I\beta\Big(C(j_\ell)+C(j_r)\Big)=0.\label{betaclass}
\end{equation}
For any real value of $\beta\neq 0$ this equation allows only the trivial solutions $j_\ell=j_r=0$, furthermore for $\beta=0$ we would have $j=j_r=j_\ell$ though, but for this case we have an unacceptable divergence in the classical action \eref{Hactn}. Consequently, using \emph{finite} dimensional representations there seems to be no obvious way to implement the reality conditions in quantum theory. Which should be understood as one of the major stumbling blocks preventing us to formulate loop quantum gravity in terms of chiral representations of $SL(2,\mathbb{C})$. 

Luckily, there is an argument against this reasoning. Instead of searching for elements $\Psi\in\mathrm{Cyl}$ being annihilated by the scalar constraint $\boldsymbol{C}\Psi=0$ one may try to implement this part of the reality conditions \emph{within} the spinfoam amplitude, that is \qq{weakly}. In fact, we search for boundary states $\Psi\in\mathrm{Cyl}$ for which
\begin{equation}
\eqalign{
\mathsf{A}_{\mathrm{vertex}}\big[\boldsymbol{C}[f]\Psi\big]&\equiv\left(\Omega\big|\boldsymbol{C}[f] \Psi\big\rangle\right.=\\
&=\int_{\mathcal{A}_{SL(2,\mathbb{C})}}\!\!\!\!\!\mathcal{D}[A]\overline{\Omega[A]}\big(\boldsymbol{C}[f]\Psi\big)[A]=0}
\end{equation}
is satisfied. Choosing a suitable regularisation\footnote[7]{Introduced for the $q$-deformed $SU(2)$ group in e.g. \cite{qdef2, qdef3}, but easily adaptable to the case of $SL(2,\mathbb{C})$.} of the squared momenta  in terms of \qq{grasping operators} the respective eigenvalues get $q$-deformed:
\numparts
\begin{eqnarray}
j_\ell(j_\ell+1) & \stackrel{q-\mathrm{def}}{\longrightarrow}& C_z(j_\ell)=\frac{[2j_\ell+2]_zj_\ell^2}{[2j_\ell]_z\mathrm{ch}(z)},\\
j_r(j_r+1)& \stackrel{q-\mathrm{def}}{\longrightarrow}& C_{\bar{z}}(j_r)=\frac{[2j_r+2]_{\bar{z}}j_r^2}{[2j_r]_{\bar{z}}\mathrm{ch}(\bar{z})},\
\end{eqnarray}
\endnumparts
where for any $a\in\mathbb{N}_0$ and $z\in\mathbb{C}$ the quantum integers are defined in the usual way, that is in terms of the hyperbolic sine function:
\begin{equation}
[a]_z=\frac{\mathrm{sh}(az)}{\mathrm{sh}(z)}\stackrel{|az|\ll 1}{\approx} a+\frac{a(a^2-1)}{6}z^2.
\end{equation}
Therefore the constraint equation \eref{betaclass} would be replaced by its $q$-deformation
\begin{equation}
\eqalign{
(\beta^2-1)&\Big(C_z(j_\ell)-C_{\bar{z}}(j_r)\Big)-2\I\beta\Big(C_z(j_\ell)+C_{\bar{z}}(j_r)\Big)=0.}\label{betadef}
\end{equation}
If we choose, e.g. $j_\ell=j_r=j$ equation \eref{betadef} further simplifies
\begin{equation}
(\beta^2-1)\mathfrak{Im}\big[C_z(j)\big]-2\beta\mathfrak{Re}\big[C_z(j)\big]=0.\label{betadef2}
\end{equation}
And this, quite contrary to the case of equation \eref{betaclass}, does have non-trivial solutions, e.g.:
\begin{equation}
\mbox{for:}\;j=\frac{1}{2},\,n=5:\beta\approx 0.55374\dots,\label{bvalue}
\end{equation}
which has, of course, been computed numerically. Furthermore equation \eref{betadef2} is easily shown to be anti-symmetric in $\beta$, which immediately implies that for any $j$ and $n$ the singular value $\beta=0$ will always solve this constraint. 

The example given in \eref{bvalue} is certainly not thought to promote particular values of $n,j$ and $\beta$ for any concrete model of quantum gravity. It is just intended to show a logical possibility---to impose and actually solve the reality conditions while using finite dimensional representations of a certain $q$-deformed $SL(2,\mathbb{C})$ group. Something which seems impossible to achieve for the \qq{classical} case (i.e. in the limit $z\rightarrow 0$).
\section{Summary and open problems}
First, we have discussed the canonical analysis of the Holst action in presence of a cosmological constant. Using complex Ashtekar variables we observed that the $SL(2,\mathbb{C})$ Kodama state formally solves all first class constraints. The level, i.e. the coupling constant appearing in the exponential in front of the Chern--Simons integral turns out to be complex, quantised and related to \emph{both} the cosmological constant \emph{and} the Barbero--Immirzi parameter \eref{lambdadef}. We then proposed to use the Kodama state in order to define a spinfoam amplitude in presence of a cosmological constant.

In section \ref{sect3} we have defined the $SL(2,\mathbb{C})$ spinnetwork states at the boundary surrounding the vertex. Each link is coloured by finite dimensional representations of the group. The vector space of boundary spinnetwork states fails to be a Hilbert space, but bears a representation of the holonomy flux algebra and can naturally be equipped with an indefinite, but non-degenerate \qq{inner product}. We encountered states of positive, negative and vanishing norm.

In section \ref{sect4}, we have computed the skein relations needed to calculate the Chern--Simons expectation value of non-intersecting loops coloured by the two representations of lowest spin. Section \ref{sect5} was more conceptual in nature. A conjecture on the relation between $SL(2,\mathbb{C})$ Chern--Simons theory and a certain \qq{chiral} deformation of the group has been presented. Next, we made a comment on the implementation of the reality conditions: We showed that on the space of the $SL(2,\mathbb{C})$ boundary spinnetworks considered, there is a serious problem. The reality conditions cannot obviously be solved. But for the $q$-deformed equations this argument no longer holds, and therefore the general framework of quantum groups may well open a way to stick with finite dimensional representations of the $SL(2,\mathbb{C})$ group.

A vast number of problems remain open, the most crucial of which are given in the following.

\emph{(i. Face amplitude.)} We have not commented on the face amplitude, needed to define the partition function \eref{spnfmmdls}. Several different choices, including
\begin{equation}
\mathsf{A}_{\mathrm{face}}(j_\ell,j_r)=(2j_\ell+1)(2j_r+1).
\end{equation}
or even a $q$-deformation thereof seem plausible to us, e.g. the former advocated by analogy with $SO(4)$ $BF$-theory and the character decomposition \eref{iddecomp} of the identity.

\emph{(ii. Regularisation for intersecting knots.)} The defining differential equation for the Chern--Simons expectation value gets singular at the intersection points making a regularisation scheme necessary. Replacing $SL(2,\mathbb{C})$ intertwiners by their respective quantum analogs may well achieve a natural way to do this. For this mechanism to work one first needs to establish the connection between $SL(2,\mathbb{C})$ Chern--Simons theory and a certain \qq{chiral} deformation of the group, a relation which has been conjectured during the last section, but still needs to be proven.

\emph{(iii. Solution for all reality conditions.)} In the last section we sketched the implementation of what we had called \qq{scalar} part \eref{scalcons} of the reality conditions \eref{realcond}. In order to legitimately claim to solve all of the reality conditions increasingly more work needs to be done. In fact we expect that the remaining parts of the reality conditions \eref{realcond} impose restrictions on the intertwiners at the nodes---a problem going to be attacked within our next paper.

\emph{(iv. Torsion.)} There is another reality condition not yet discussed so much in spinfoam gravity. Torsion must vanish. At this stage we can say only very little about the implementation of this constraint, and must leave a complete analysis undone.

\emph{(v. Kinematical operators.)} One of the most exciting tasks to consider concerns the usual \cite{rovelli, thiemann} geometrical operators of loop quantum gravity, i.e. the question whether the discrete eigenvalues of area and volume \cite{rovelli} can naturally be recovered within our formalism. 

\emph{(vi. Infinite dimensional representations.)} The present Lorentzian spinfoam models, e.g. the one defined by the EPRL amplitude \cite{lorentzvertam, LQGvertexfinite} are all formulated in terms of infinite dimensional representations \cite{gelminshap} of the Lorentz group. Having used finite (chiral) representations instead, we immediately have to face the question of the relation between them two. 
Is it possible to understand one model in the kinematical framework of the other, or are they fundamentally incommensurable? 

\emph{(vii. Relation to Euclidean models.)}
In a paper \cite{muxinvert} appearing parallel to this one Muxin Han studies the Euclidean version of four dimensional gravity in presence of a cosmological constant. He arrives at a model sharing key features with the one studied here, studies the asymptotic behaviour of the amplitude considered, and actually proves that the model converges to the Euclidean Regge action equipped with a cosmological constant.
\section*{Acknowledgements} 
Let me express my gratitude and thanks to my friends and colleagues involved in the creation of this article. I wish to thank my supervisor Carlo Rovelli for constant advise and personal support, spreading motivation and joy in physics. 
I am indebted to Eugenio Bianchi and Muxin Han for endless discussion, friendship and help. They always had time for my problems, be they simple, be they hard to go through.
\section*{References}

%\bibliography{C:/Users/wmw/Documents/Bibliographie/LQG}
%\bibliographystyle{C:/Users/wmw/Documents/Bibliographie/naturemagv}
%\bibliographystyle{amsalpha}

\end{document}